\documentstyle{ubcthesis}
\newcommand{\Vlasov}{Vlasov}
\newcommand{\resetcounters}{ 
	\setcounter{equation}{0} 
	\setcounter{figure}{0} 
	\setcounter{table}{0}   }       
\input epsf  
\epsfverbosetrue
\begin{document}
\bibliographystyle{unsrt}

   \thesistitle{\bf Critical Collapse 
of Collisionless Matter in Spherical Symmetry}
\author{Ignacio (I\~naki) Olabarrieta}
\previousdegrees{Licenciado en Ciencias, 
Universidad del Pa\' {\i}s Vasco, 1998}
\degreetitle{Master of Science}
\department{Department of Physics and Astronomy}
\signaturelines{2}

\copyrightyear{September 2000}

\titlepage
\authorizepage
\begin{center}
 {\Huge \bf Abstract}     
\end{center}
\addcontentsline{toc}{chapter}{Abstract}
\vspace{2ex} 
We perform a numerical study of the critical regime for the 
general relativistic collapse of collisionless matter in spherical symmetry. 
The evolution of the matter is given by the \Vlasov\ equation
(or Boltzmann equation) and the geometry by Einstein's equations. 
This system of coupled differential equations is solved using a particle-mesh (PM) 
method. This method approximates the distribution function which describes 
the matter in phase space with a set of particles moving along the characteristics
of the \Vlasov\ equation. The individual particles are allowed to have 
angular momentum different from zero but the total angular momentum 
has to be zero to retain spherical symmetry. 

In accord wih previous work by Rein, Rendall and Schaeffer, our results 
give some indications that the critical behaivour in this model is of 
Type I (the smallest black hole in each family has a finite mass).
For the families of initial data that we have studied it seems that in
the critical regime the solution is a static spacetime with non-zero radial
momentum for the individual particles. 
We have also found evidence for scaling laws for the time that the critical solutions spend in the critical regime.
\newpage
\addcontentsline{toc}{chapter}{Table of Contents}
\tableofcontents
\listoftables
\listoffigures
\newpage
\begin{center}
 {\Huge \bf Acknowledgements}    
\end{center}
\addcontentsline{toc}{chapter}{Acknowledgements}
\vspace{2ex}
I want to thank the rest of the graduate students under the supervision of Matt Choptuik, 
especially Jason Ventrella, with whom I have had many useful discussions. 
I also want to thank Bojan Losic, Noureddine Hambli, Luis Lehner and Bill Unruh who 
have been there for any question I ever had. I am grateful to all the people who have
encouraged me from Spain, my family and friends, and the people 
of the department of {\em Theoretical Physics and History of Science} in the 
University of the 
Basque Country who allowed me to work there for some time during the development of 
this thesis. Finally, my supervisor, Matt Choptuik, for keeping on asking me 
{\it where are the particles?} and for being a friend in addition to a supervisor.   

%

\resetcounters

\chapter{Introduction}
\label{intro}

\section{Gravitation and Critical Phenomena}
\label{beginning1}
Einstein's theory of gravitation connects the geometry of 
spacetime to its matter content 
via the field equations\footnote{We will use units where $c=1$ and 
$G=1$ ($c$ being the speed of light, and $G$ Newton's 
gravitational constant)}:
\begin{equation}
\label{Einstein}
G_{ab}=8 \pi T_{ab}.
\end{equation}  
Here $G_{ab}$ is the Einstein tensor, whose coefficients are complicated 
functions of the metric, $g_{ab}$, and its first and second derivatives.
$T_{ab}$ is the stress energy tensor, which depends on the matter, and also,
in general, on the metric. 

In a given coordinate system, these equations are non-linear partial differential equations for the
metric coefficients and can give rise to very interesting solutions for the 
spacetime.
Some of the most interesting phenomena occur at the
threshold of black hole formation as was first discovered 
numerically by Choptuik \cite{Choptuik:1993},
who studied the collapse of a massless scalar field
in spherical symmetry.

For initial data characterized by one parameter $p$ 
(the maximum amplitude of the scalar field for example), Choptuik 
found a critical value $p^*$ such that for 
values of $p$, $p>p^*$, the evolution gave rise to the
formation of a black hole, while for $p<p^*$ the scalar field dispersed to
infinity. Near $p=p^*$ the space-time approached a universal solution,
independent of the particular choice of the initial shape of the pulse
(e.g. gaussian, tanh, etc). 
It was found that the near-critical dynamics was characterized by self-similar 
oscillations with scaling factor $e^\Delta$. 

Choptuik also found that the masses of the 
black hole formed obeyed a scaling law:
\begin{equation}
M_{BH} \propto |p-p^*|^\gamma
\end{equation} 
where $\gamma$ was  a universal exponent, again, universal in the sense of 
being independent of the 
choice of the family of initial data. This equation shows that, in principle,
one can form black holes with arbitrarily small masses in the
model. The transition to 
black hole formation in this case is thus continuous in the mass of the 
black hole. 
In analogy with the critical behavior 
in statistical mechanics this was called a Type II critical solution.  

A lot of work has been done in the last few years trying to find the critical
solutions for different types of matter. 
Type I transitions (i.e. transitions where the mass of the smaller black hole
is finite) have been also found \cite{ym}. In this case, the
solutions that have been found are either static or periodic \cite{Gundlach}. 
Here it was found that the time that the solution stays in the critical 
regime scales like:
\begin{equation}
\label{scal}
\tau \sim -\sigma \ln{|p-p^*|} 
\end{equation} 
where, again, $\sigma$ is a universal constant\footnote{Note that this 
constant depends on the units used.}.


The process of tuning $p$ to $p^*$ can be understood as choosing 
initial data such that at the threshold of black hole formation, we 
tune-out any growing component. Since we can {\em tune away} the 
growing components using only one parameter, the solution apparently has 
exactly one unstable mode. Linear perturbations of the critical solutions
 \cite{Gundlach} agree with that observation. 

%
%

In this thesis we study the critical regime for collisionless matter in 
spherical symmetry. If we think of the matter as being composed of individual particles,
allowing these particles to have angular momentum 
different from zero (keeping the net angular momentum equal to zero to remain in spherical symmetry) 
can give us some insight concerning the importance of
angular momentum in critical phenomena, without needing to go to spacetimes 
with less symmetry.        

\section{Collisionless Matter (A little bit of history)}
\label{beginning2}
Collisionless matter (or dust) has been widely studied in general relativity.
Einstein \cite{Einstein:1939} 
used this type of matter to investigate
the physical significance of the singularity at $r=2\,M$ for the Schwarzschild solution. 
He studied clusters of particles rotating in circular orbits 
(and spherical symmetry) and proved that for clusters in equilibrium 
the Schwarzschild mass function cannot reach $M(r)=r/2$ 
($r$ being the areal coordinate), 
therefore these  configurations
cannot have an event horizon nor a singularity. 
He postulated that this result also would hold for systems not in equilibrium and 
concluded that in ``physical reality" there is no configuration of matter such that
$M(r)=r/2$ (i.e. black holes cannot form). 
Although he came to the wrong conclusion, he studied one of the 
first models for relativistic stars.

In the case when the angular momentum of each particle is zero,
and the spacetime is spherically symmetric, 
Oppenheimer and Snyder \cite{Oppemheimer:1939} showed that a homogeneous 
distribution of this kind of matter generally forms a singularity.
Using comoving coordinates they found the solution for 
the spacetime and deduced that singularity formation would happen 
in finite proper time (at least from the point of view of the comoving
observers).
Tolman \cite{Tolman:1934} and Bondi \cite{Bondi:1947} 
generalized this result for non-homogeneous configurations, 
restricting to the case where no particle overtakes any other during 
the evolution.

Eardley and Smarr \cite{Eardley:1979}
considered the same model as Tolman and Bondi 
and studied the existence of maximal ($K=0$) and constant mean curvature ($K=K_o$)
slicing conditions.
Shapiro and Teukolsky
\cite{Shapiro:1985a}, \cite{Shapiro:1985b}, \cite{Shapiro:1985c},
\cite{Shapiro:1986}, \cite{Shapiro:1989} have studied this system in detail
in the context of  models of stellar clusters. They developed a code very similar 
to ours to study properties of relativistic clusters, such as stability and 
collapse to a black hole. They also extended the code to treat more general spacetimes 
, i.e. with axial symmetry, \cite{Shapiro:1991} to study naked singularities. 

Rendall \cite{Rendall:1993}, \cite{Rendall:1996}, and Rein \cite{Rein:1993} have a series 
of papers addressing general properties for the Einstein-Vlasov system 
(the Cauchy problem, 
existence of static solutions
in spherical symmetry, existence of general asymptotically flat solutions, etc...).


More recently, Rendall, Rein and Schaeffer \cite{Rendall:1998} 
have published the first study of critical behavior (black-hole threshold 
behavior) for this type of matter. Indeed, in this last paper Rendall 
{\em et al} argue that the critical solution
in the case of collisionless matter is generically Type I. Our results agree with
theirs in that we also find evidence for \break Type I transitions for 
certain initial data families which we have studied in this thesis. In addition to that result,
we argue here that the critical solutions in this model are static, and we present evidence for scaling laws of the form (\ref{scal}). Finally we have preliminary investigations aimed at determining whether the critical solution is unique
(universality).

\resetcounters

\chapter{The Equations}
\label{ch1}

The dynamical state of dust can be described by a distribution function, $f$:
\begin{equation}
\label{distri}
f(x^a,p_a)=dN/dV_p
\end{equation}
where $N$ is the number of particles and $V_p$ is the volume in phase space.
In our case, and since the matter is collisionless, the volume in phase space 
is a conserved quantity during the evolution of the system (Liouville theorem). 
This implies that the distribution function is also a conserved quantity:
\begin{equation}
\label{vlasov}
\frac{df}{d\tau}=0
\end{equation}   
This is the collisionless Boltzmann or \Vlasov\ equation.  
This equation plus Einstein's equations (\ref{Einstein}),
all restricted to spherical symmetry, i.e.:
\begin{equation}
f(t,x^i,p_i)=f(t,R \, x^i,R\, p_i) \quad {\mathrm{with}}\quad R \in SO(3) \quad i=1,2,3
\end{equation}
form the system of equations that we want to solve
numerically.

Numerical solutions of PDEs generally involves discretization of the continuum 
problem, and here we can consider at least two approaches: 
1) Consider a set of particles which approximates the distribution function at
the initial time and evolve the particles along the characteristics of equation (\ref{vlasov}). 
The 
geometry is computed at each time step using the energy densities 
derived from the positions and velocities of the particles.
2) Discretize the \Vlasov\ equation for the distribution function in phase space  
and solve for it directly in phase space. In this case the energy 
densities, required in the update of the geometry, are calculated 
by integrating the distribution function over momentum space.


In the particle approach, we have the problem that a given set of particles
is just one of the infinite number of possible realizations for given density and 
velocity profiles, therefore we introduce statistical errors.
In the second case
we have to solve the \Vlasov\ equation, a partial differential 
equation in phase space. 
Although
we have taken the particle approach due to the 
simplicity of the evolution equations,
we include the equations for the direct integration in Appendix B.

As with any problem in ``3+1" (or ADM) numerical relativity, we want to be able to specify
initial data on a spatial hypersurface  and then evolve these data in time. 
To do this, we need to split the 
equations (\ref{Einstein})  
into a set of constraint equations (equations that
must be satisfied at each instant of time) and dynamical or
evolution equations (equations that tell us how to evolve the geometric 
quantities in time). In the most general case, we will have four 
constraint and six second-order-in-time evolution equations.

In order to do this splitting, we will make use of the $3+1$ formalism due to
Arnowitt, Deser and Misner (ADM), for a review of this formalism see \cite{MTW}. We have
followed a summary by Choptuik \cite{3+1} which is itself based on a 
review by York \cite{YorkSmarr}.

In the  remainder of this chapter, we derive the system of equations that we solve.
We start with a summary of the $3+1$ formalism and its restriction to the spherical
symmetric case, and a definition and description of the maximal-areal coordinate system.    
In section 2.2 we explain how the Einstein's equations are coupled to the matter, and in 2.3 
we explain the particle evolution equations in our coordinates. The chapter ends with some calculations 
of conserved quantities.  

\section{The 3+1 Formalism of General Relativity.}
We consider a spacetime manifold with metric $g_{ab}$ defined on it, 
and assume that we can slice the manifold into spacelike hypersurfaces, defined
by $t=$~constant, where $t$ is our time coordinate.  At least locally then,
the space-like surfaces can be  described in terms of a closed one-form, 
$\Omega_a$\footnote{Indexes in this section are abstract indexes as in \cite{Wald}},
which is just the gradient of $t$
\begin{equation}
\Omega_a=d t=\nabla_a t.
\end{equation}
The norm of this one-form is:
\begin{equation}
g^{ab}\Omega_a \Omega_b = - \alpha^{-2},
\end{equation}
where $\alpha$ is a scalar function commonly known as the lapse function.
As long as our slices {\em are} spacelike, $\alpha$ will be strictly positive.
We introduce a normalized one-form, $n_a$, defined by
\begin{equation}
n_a=-\alpha \Omega_a = -\alpha \nabla_a t.
\end{equation}
The contravariant vector, $n^a=g^{ab}n_b$, which is the future-directed normal 
to the surfaces, can be viewed as the four
velocity field of observers moving orthogonally to the slices. 

In this section, again following York, we will describe the 3+1 
split in terms of 4-dimensional tensors.  At the same time, 
we will need to decompose various tensors (including the Einstein
and stress energy tensors appearing in (\ref{Einstein})) into pieces
parallel to $n^a$ (``timelike'' pieces) and pieces orthogonal to
$n^a$ (``spacelike'' pieces). 
Thus we define
\begin{equation}
W^n=-W^an_a,
\end{equation}
where the sign convention is again York's, while for covariant
vectors we define
\begin{equation}
W_n=W_a n^a.
\end{equation}
For the projection onto the hypersurfaces we use the projection operator:
\begin{equation}
\gamma^a{}_b=\delta^a{}_b+n^an_b
\end{equation}
The projection of the space-time metric to the
hypersurfaces is:
\begin{equation}
\gamma_{ab}=\gamma^c{}_a \gamma^d{}_b\, g_{cd}=g_{ab}+n_a\,n_b
\end{equation}
and is, in fact, the metric induced on the hypersurfaces by the 3+1 splitting.
(Note that $\gamma^{ab}$ is not necessarily the inverse of $\gamma_{ab}$, as follows 
immediately from the definition of $\gamma^{a}{}_b$, and in this 
section we raise and lower all indexes with the four dimensional metric $g^{ab}$ and 
its inverse).
We can now introduce the natural derivative operator on the space-like surfaces, 
by projection of the four dimensional covariant derivative:
\begin{equation}
D_a= \gamma^b{}_a \, \nabla_b.
\end{equation}
It is easily shown that this derivative is compatible with the spatial metric:
\begin{equation}
D_a \gamma_{bc} = 0.
\end{equation}

Having defined a derivative operator on the hypersurfaces, we can compute 
the associated Riemann tensor ${{{}^3 {\cal R}}_{abc}}^d$
which measures the intrinsic curvature of the hypersurface.  For example, for a 
{\em spatial} one form $\omega_a$ ($\omega_a n^a = 0$) we have

\begin{equation}
   \left( D_a D_b - D_b D_a \right) \omega_c = {{{}^3 {\cal R}}_{abc}}^d \omega_d.
\end{equation}
We can then also define the spatial Ricci tensor,
${{}^3 {\cal R}}_{ac} = {{{}^3 {\cal R}}_{abc}}^b$ and the spatial Ricci
scalar, ${}^3 {\cal R} = {{{}^3 {\cal R}}^a}_a$.
\\
Finally, we define the extrinsic curvature tensor, $K_{ab}$:
\begin{equation}
K_{ab}=-D_a n_b=\nabla_a n_b - n_a a_b,
\end{equation}
where $a_b = n^a D_a n_b$ is the dual to the four acceleration field of the observers 
moving with the slicing.
\\
The geometry of the space-time is completely defined by the
spatial metric, $\gamma_{ab}$, which describes the geometry of 
each slice, 
and the extrinsic curvature tensor which tells us how 
each slice is embedded in the four dimensional space-time.

We can write Einstein's equations in terms of the tensors 
defined above, as well as the following
projections of the stress energy tensor:
\begin{eqnarray}
\label{srho}
\rho    &=& T^{ab}n_a n_b\\
j^a     &=& \gamma^a{}_b(T^{bc} n_c) \\
\label{sstress}
S^{ab}  &=& \gamma^a{}_c \gamma^b{}_d T^{cd}
\end{eqnarray}
These are interpreted as the local energy density,
momentum density and spatial stress tensor, respectively, for an observer 
moving orthogonally to the space-like hypersurfaces. 
\\
The Hamiltonian constraint is found by contracting Einstein's
equations with $n^a$ twice,
$G_{ab}n^a n^b=T_{ab} n^a n^b$, yielding
\begin{equation}
{{}^3 {\cal R}} + K^2 -{K^a}_b{K^b}_a = 16 \pi \rho
\end{equation}
where $K = g^{ab}\,K_{ab} = \gamma^{ab}\,K_{ab}$ is the trace of the extrinsic curvature and ${{}^3 \cal R}$ is the spatial Ricci
scalar defined previously.
\\
If we contract Einstein's equations once with $n^a$, and then project the 
resulting vector onto the
hypersurface, $\gamma^a{}_c \, G^{cb} \, n_b= \gamma^a{}_c  T^{cb}  n_b$, we get the momentum constraint:
\begin{equation}
D_b K^{ab} - D^a K = 8 \pi j^a.
\end{equation}
These two constraint equations involve only spatial tensors and spatial
derivatives of spatial tensors, and must be satisfied on each spacelike 
slice, including the initial slice.
\\

In order to derive the evolution equations we use the Lie derivative
 along the vector field, $N^a$:
\begin{equation}
N^a=\alpha n^a + \beta^a=d /dt
\end{equation}
where $\beta^a$ is an arbitrary spatial vector field, commonly known 
as the {\em shift vector}.
Note that $N^a$ satisfies:
\begin{equation}
N^a \Omega_a = 1
\end{equation}
It can be shown that Lie differentiation along $N^a$ commutes 
with the projection operator. 
We can now write the rest of equations (\ref{Einstein}) as two different sets:
(1) the evolution of the spatial metric:
\begin{equation}
{\cal L}_t \gamma_{ab} = -2 \alpha K_{ab} + {\cal L}_\beta \gamma_{ab}
\end{equation}
which can also be viewed as a definition of $K_{ab}$,
and (2) the evolution equations for the extrinsic curvature:
\begin{equation}
\label{evolK}
{\cal L}_t {K^a}_b = {\cal L}_\beta {K^a}_b -D^aD_b \alpha +\\
\alpha{{}^3{\cal R}^a}_b + K{K^a}_b + 8 \pi 
\left(\frac{1}{2} {P^a}_b \left(S-\rho \right)-{S^a}_b \right)
\end{equation}
where $S$ is the trace of tensor given by equation (\ref{sstress}).
\subsection{3+1 Formalism in Spherical Symmetry}
We now restrict our attention to
spherical symmetry.  In contrast to the previous 
section, where our tensor expressions involved
abstract 4-dimensional indices, we will be 
mostly concerned with the {\em components} of 
specific 3-tensors in this section.  Thus, Latin
indices $i, j, k, \cdots$ range over the spatial
values $1, 2$ and $3$, and all indices of the spatial tensors are raised 
and lowered with $\gamma^{ij}$ and 
$\gamma_{ij}$ respectively, where $\gamma^{ij}\,\gamma_{jk}=\delta^i{}_k$.

The most general 3-metric in spherical 
symmetry can be written as:
\begin{equation}
{}^3d{s}^2=a^2\left(t,r\right)\,dr^2+r^2\,b^2\left(t,r \right) \,d\Omega^2
\label{Spherical}
\end{equation}
where $d\Omega^2$ is the metric on the unit sphere, $d\Omega^2=d\theta^2 + \sin^2{\theta}\,d\phi^2$.
From the 3-metric, we can compute the associated connection coefficients or Christoffel symbols:
\begin{equation}
\label{Christoffel}
\Gamma^i{}_{jk}=\frac{1}{2}\gamma^{in}
\left(\gamma_{nj,k}+\gamma_{nk,j}-\gamma_{jk,n} \right)
\end{equation}
In our coordinates, the non-zero connection coefficients are:
\begin{equation}
\begin{array}{c}
\begin{array}{lll}
\Gamma^r{}_{rr} = a^{\prime}/a &\quad \quad 
\Gamma^r{}_{\theta \theta}=-\left(\left(r^2b^2\right)^{\prime}\right)/
\left(2a^2\right) \quad \quad&
\Gamma^r{}_{\phi \phi}=-\sin^2{\theta}\left(r^2b^2\right)^{\prime}/
\left(2a^2\right)\\
\end{array}\\
\\
\begin{array}{ll}
\Gamma^\theta{}_{r \theta}=\left(
\left(r^2b^2 \right) ^\prime \right)/\left(2r^2b^2\right)&\quad \quad \quad
\Gamma^\theta{}_{\phi \phi}=-\sin{\theta}\cos{\theta}\\
\\
\Gamma^\phi{}_{r \phi}=\left(r^2b^2 \right)^\prime/ \left(2r^2b^2 \right)&\quad \quad \quad 
\Gamma^\phi{}_{\theta \phi}=\cot{\theta}
\end{array}
\end{array}
\end{equation}
We also need to compute the components of the Ricci tensor:
\begin{equation}
{}^3R_{ij}=\Gamma^n{}_{ij,n}-\Gamma^n{}_{in,j}+\Gamma^n{}_{nm}\,\,\Gamma^m{}_{ij}-\Gamma^n{}_{jm}\,\,\Gamma^m{}_{in}
\end{equation}
from which we find the non-zero components:
\begin{equation}
\begin{array}{l}
{}^3R_{rr}=-\left[\left(r^2b^2\right)^\prime/\left(r^2b^2\right)\right]^\prime+
        a^\prime \left(r^2b^2\right)^\prime/\left(a r^2b^2\right)-
        \frac{1}{2}\left[ \left(r^2b^2\right)^\prime/\left(r^2b^2\right)\right]^2\\
\\
{}^3R_{\theta \theta}=-\left[\left(r^2b^2\right)^\prime/\left(2a^2\right)\right]^\prime+1-
                  \left[a^\prime\left(r^2b^2\right)^\prime/\left(2a^3\right)\right]\\
\\
{}^3R_{\phi \phi}=\sin^2{\theta}R_{\theta \theta}
\end{array}
\end{equation}
The mixed components are given by
\begin{equation}
\begin{array}{c}
{}^3R^r{}_r=-2\left[ \left(rb\right)^\prime/a\right]^\prime/\left(arb\right)\\
\\
{}^3R^\theta{}_\theta=R^\phi{}_\phi=\left[ a - \left(rb\left(rb\right)^\prime/a\right)^\prime\right]
/\left(a\left(rb\right)^2\right).
\end{array}
\end{equation}
Another quantity which we need is ${\alpha_{|i}}^{|j}$, where ``$\vert$'' denotes
covariant differentiation with respect to the 3-metric.
\begin{equation}
{\alpha_{|i}}^{|j}=\gamma^{jk}\left(\alpha_{,ik}-\Gamma^n{}_{ik}\,\alpha_{,jn}\right)
\end{equation}
which has the following non-zero components:
\begin{equation}
\begin{array}{lcr}
{\alpha_{|r}}^{|r}=\left( \alpha^\prime/a\right)^\prime/a
\quad \ &
{\alpha_{|\theta}}^{|\theta}=\alpha^\prime\left(r^2b^2\right)^\prime/
\left(2r^2 a^2 b^2\right)
\quad \ &
{\alpha_{|\phi}}^{|\phi}=\left(rb\right)^\prime\alpha^\prime/\left(rba^2\right)
\end{array}
\end{equation}
Finally, the Ricci scalar, ${}^3R=\gamma^{ij}R_{ij}$, is given by:
\begin{equation}
{}^3 R=\frac{2}{arb}\left[-2\left(\frac{\left(rb\right)^\prime}{a}\right)^\prime +
  \frac{a}{rb}\left(1-\frac{{\left(rb\right)^\prime}^2}{a^2}\right) \right]
\end{equation}
At this point we have calculated all the geometric objects intrinsic to the spatial slices 
that we need.
Now, we embed these slices into the most general 
spherically symmetric {\em spacetime} whose metric can be written:
\begin{equation}
ds^2=\left(-\alpha^2+a^2\beta^2\right)dt^2+2a^2\beta dtdr+a^2dr^2+r^2b^2d\Omega^2
\label{SphericalFour}
\end{equation}
where $\alpha$, $\beta$, $a$ and $b$ are all functions of 
$r$ and $t$.  Note that due to the spherical symmetry, 
the shift vector has only a radial component; 
$\beta^i = \left(\beta,0,0\right)$, so $\beta$ will be
called the shift function.
In these coordinates $n_\mu=\left(-\alpha,0,0,0\right)$ and
$n^\mu=\left(1/\alpha,-\beta/\alpha,0,0\right)$.

The extrinsic curvature can be written in terms of the 4-metric coefficients:
\begin{equation}
K_{ij}=-\frac{1}{2\alpha}\left(\frac{\partial \gamma_{ij}}{\partial t}-\gamma_{ij,k}\beta^k -
               \gamma_{ik}\beta^k{}_{,j}-\gamma_{kj}\beta^k{}_{,i}\right)
\end{equation}
From this we compute the non-vanishing components for the current spherically
symmetric case:
\begin{eqnarray}
K_{rr}&=&\frac{a}{\alpha}\left(\left(a \beta\right)^\prime- \dot a \right) \\
K_{\theta \theta}&=&\frac{rb}{\alpha}\left(\beta\left(rb\right)^\prime-r \dot b \right) \\
K_{\phi \phi}&=&\sin^2{\theta}K_{\theta \theta}
\end{eqnarray}
The corresponding components with mixed indexes are:
\begin{eqnarray}
K^r{}_r&=&\frac{1}{a \alpha}\left(\left(a \beta\right)^\prime- \dot a \right) \\
\label{ktt}
K^\theta{}_\theta&=&\frac{1}{rb\alpha}\left(\beta\left(rb\right)^\prime-r \dot b\right) \\
K^ \phi{}_\phi&=&K^\theta{}_\theta
\end{eqnarray}
and the trace of the extrinsic curvature is
\begin{equation}
K=\frac{1}{a \alpha}\left( \left(a \beta\right)^\prime-\dot a \right)+
\frac{2}{\alpha}\left( \beta \frac{\left(rb\right)^\prime}{rb}-\frac{\dot b}{b}\right)
\end{equation}
Finally, the only non-zero component of $D_i K \equiv K_{\vert i}$ is 
\[
K_{|r}=K_{,r}=-\frac{1}{a \alpha}\left( \frac{a^\prime}{a}+\frac{\alpha^\prime}{\alpha}\right)
              \left(a \beta^\prime -
              \dot a \right) +
              \frac{1}{a \alpha}\left((a \beta)^{\prime \prime}-\dot a^\prime \right) -
             \frac{2}{\alpha^2}\alpha^\prime\left(\beta\frac{(rb)^\prime}{rb}-1\right)
\]

\subsection{Maximal Areal Coordinate System.}
Einstein's equations allow for coordinate freedom that we need to fix.
We have chosen maximal-areal coordinates (defined below), but have also
included the equations for the polar-areal system in Appendix C.
The main advantage of maximal-areal coordinates is that, in contrast 
to the polar-areal case, the slices used can penetrate apparent 
horizons.  As the name suggests, in the maximal-areal coordinate
system,
the radial coordinate is areal, so that
the proper area of 2-spheres with radius $r$ is $4\pi r^2$.
In terms of the general spherically-symmetric 4-metric (\ref{SphericalFour}),
this  means that $b\left(r,t\right) \equiv 1$.  The time coordinate is fixed by
demanding that the 3-slices be {\em maximal}, i.e. that $K\left(r,t\right) = 0$.
This leads to a condition on the lapse function $\alpha\left(r,t\right)$ (the 
so-called slicing condition), which must be satisfied at each instant
of time.

Thus, in maximal-areal coordinates the 4-metric (\ref{SphericalFour})
takes the form:
\begin{equation}
\label{metric}
ds^2=\left(-\alpha^2+a^2\beta^2\right)dt^2+
     2a^2\beta dtdr+a^2dr^2+r^2d\Omega^2
\end{equation}
Using the expressions we derived in the previous section, we compute the relevant geometrical 
quantities we need in order to write Einstein's equations in terms of these metric coefficients:
\begin{equation}
\begin{array}{c}
R=2\left[-2\left(1/a\right)^\prime+
  a/r\left(a-1/a^2\right)\right]/(a\,r)\\
\\
\begin{array}{ll}
R^r{}_r=-2\left(1/a\right)^\prime/(a\,r) \quad \quad \quad \ &
R^\theta{}_\theta=R^\phi{}_\phi=
\left[a-\left(r/a\right)^\prime\right]/(a\,r^2)\\
\\
K^r{}_r=\left(\left(a\,\beta\right)^\prime - \dot a \right)/(a\,\alpha)\quad \quad \quad \ &
K^\theta{}_\theta=K^\phi{}_\phi=\beta/\left(r\, \alpha\right)\\
\\
{K^i{}_r}{}_{|i}=K^r{}_{r,r}+2\left(K^r{}_r-K^\theta{}_\theta\right)/r\quad \quad \quad \ &
K_{|r}=0\\
\\
{\alpha_{|r}}^{|r}=\left(\alpha^\prime/a\right)^\prime/a\quad \quad \quad \ &
{\alpha_{|\theta}}^{|\theta}={\alpha_{|\phi}}^{|\phi}=\alpha^\prime/\left(r\,a^2\right)
\end{array}
\end{array}
\end{equation}
We can now specialize equations (\ref{Einstein}):

{\it Hamiltonian Constraint:}
\begin{equation}
\label{geo_eqn1}
\frac{a^\prime}{a}=\frac{3}{2}a^2r{K^\theta{}_\theta}^2+4\pi r a^2\rho+\frac{1}{2r}
\left(1-a^2\right)
\end{equation}

{\it Momentum Constraint:}
\begin{equation}
{K^\theta{}_\theta}^\prime=-\frac{3}{r}K^\theta{}_\theta-4\pi j_r
\end{equation}

{\it Evolution of the 3-metric:}
\begin{eqnarray}
\dot a&=&2 \alpha a K^\theta{}_\theta+\left(a\beta\right)^\prime\\
\beta&=&\alpha r K^\theta{}_\theta
\end{eqnarray}

{\it Slicing Condition:}
\begin{equation}
\label{geo_eqn2}
\alpha^{\prime \prime}=\alpha^\prime\left( \frac{a^\prime}{a}- \frac{2}{r}\right)+
     \frac{2 \alpha}{r^2}\left( 2r \frac{a^\prime}{a}+a^2-1 \right)+4 \pi a^2 \alpha 
\left(S-3\rho \right)
\end{equation}
where in deriving the last formula, we have made use of the slicing condition 
($K=0$), and the evolution equations for the extrinsic curvature components,
 equations (\ref{evolK}). 
As we will discuss in more detail below, 
we have chosen to do a constrained evolution, which, in this case  means
that we use the constraint equations, rather than
evolution equations, to update $a$ and $K^\theta{}_\theta$.
\section{Stress-Energy Tensor}
In this section we explain how we calculate the energy densities $\rho$, $j_r$ 
and $S$ that appear in equations (\ref{geo_eqn1}-\ref{geo_eqn2}). We approximate the 
distribution function (\ref{distri}) by a set of $N$ ``spherical particles".
Since the particles only interact with each other gravitationally, we have
\begin{equation}
T^{\mu \nu}=\sum _{i=1}^N T^{\mu \nu}_i
\end{equation}
where $T^{\mu \nu}_i$ is the stress energy tensor for a single particle. 
For a point particle we can write:
\begin{equation}
\label{tmunu}
T^{\mu \nu}_i =\frac{p^\mu_i\, p^\nu_i}{m_i}
\delta(\vec{r}-\vec{r}_i(t)).
\end{equation}
Here, $p^\mu _i$ is the $\mu$ component of the 4 momentum of the $i$-th
particle, $m_i$ is its rest mass, and $\vec r_i(t)$ is the spatial position of the particle at time t.

Using equations (\ref{srho}-\ref{sstress}) for $\rho$, $S$ and $j_r$ specialized to our coordinates, we get:
\begin{eqnarray}
\left[\rho\right]_i&=&\alpha^2 \left[T^{tt}\right]_i\\
\left[S\right]_i&=&\frac{1}{a^2} \left[T_{rr}\right]_i + 
\frac{1}{r^2}\left[T_{\theta\theta}\right]_i + \frac{1}{r^2 {\mathrm sin}^2{\theta}}\left[T_{\phi\phi}\right]_i\\
\left[j_r\right] _i&=&\alpha \left[T^t{}_r\right] _i
\end{eqnarray}
Now we relax the point particle approximation and assume that each particle 
is a spherically symmetric shell of mass, uniformly distributed over a 
region $\Delta r$ of space\footnote{
Later on this $\Delta r$ will be the mesh spacing used in the finite difference 
solution of the geometrical equations.}. These shells of matter are averages of 
the shells at that point with magnitude of the angular momentum $l$ pointing 
in all posible directions.
Therefore the angular momentum of the average is zero, $\vec l=0$, but $l^2 \neq 0$.
The proper volume that each particle occupies is then:
\begin{equation}
V_i = \frac{p^t}{m_i}\,\Delta r\int \sqrt{-g}\,\, d\phi\, d\theta
    =4 \pi\, \Delta r\, \alpha\, a\, \frac{r^2_i \,\, p^t_i}{m_i}
\end{equation}
and we can approximate the delta function that appears in  
(\ref{tmunu}) by $1/V_i$.
This yields:
\begin{eqnarray}
\rho_i&=&\frac{1}{4\pi\Delta r a} \frac{\left[\bar p^t\right]_i}{r^2_i}\\
\label{sr_sa}
S_i=\left[S^r{}_r \right]_i+\left[S^a{}_a \right]_i&=&\frac{1}{4\pi\Delta r a^3 } 
    \frac{{\left[p_r\right]_i}^2}{\left[\bar p^t\right]_ir^2_i}+
    \frac{1}{4\pi\Delta r a} \frac{\left[l\right]^2_i}{r^4_i \left[\bar p^t\right]_i}\\
\left[ j_r \right] _i&=&\frac{1}{4\pi\Delta r a} \frac{\left[p_r\right]_i}{r^2_i}
\end{eqnarray}
Here $a$, $\alpha$ and $\beta$ are evaluated at $r=r_i$,  
$\left[\bar p^t\right]_i$ is defined as $\left[\bar p^t\right]_i \equiv \alpha \left[p^t\right]_i$, and 
${\left[l\right]_i}^2 \equiv {\left[p_\theta \right]_i}^2+{\left[p_\phi\right]_i}^2 / \sin^2{\theta_i}$ 
is the square of the magnitude of the angular momentum of the $i$-th particle. 
We then introduce quantities which do not explicitly
depend on the geometry (since after updating the particle positions we want to 
solve for the geometry): 
$\left[ \bar \rho \right]_i \equiv a \left[\rho \right]_i$,
$\left[\bar S^r{}_r\right]_i \equiv a^3 \left[S^r{}_r \right]_i$, $\left[ \bar S^a{}_a\right]_i = a 
\left[S^a{}_a \right]_i$ and $\left[\bar \j_r\right]_i \equiv  a \left[j_r\right]_i $.  

We interpolate the one-particle quantities to the continuum 
and sum over all the particles to find the total values:
\begin{equation}
\label{inter}
\bar f=\sum_{i=1}^{N_p} \bar f_i W(r-r_i)
\end{equation}
where $\bar f_i$ is any of the barred quantities defined above for each particle,
$\bar f$ is the corresponding quantity in the continuum case, and $W(r-r_i)$ is an 
interpolation function (see section \ref{interpolation} for further explanation of how
we define this function). 
Having defined (\ref{inter}) we can now write equations (\ref{geo_eqn1}-\ref{geo_eqn2}) as 
\begin{eqnarray}
\label{field1}
\frac{a'}{a}&=&\frac{1-a^2}{2r}+\frac{3}{2}ra^2 {K^\theta{}_\theta}^2 + 4\pi
a r \bar \rho\\
 \label{field2}
{K^\theta{}_\theta}'&=&-\frac{3}{r}K^\theta{}_\theta-4\pi \frac{\bar \j_r}{a}\\
\alpha''&=&\alpha'\left(\frac{a'}{a}-\frac{2}{r}\right)+\frac{2\alpha}{r^2}
\left(a^2-1+2r\frac{a'}{a}\right)+4\pi
a \alpha \left(\frac{\bar S^r{}_r}{a^2}+\bar S^a{}_a-3 \bar \rho \right)\\
\label{field4}
\beta&=&\alpha r K^\theta _\theta
\end{eqnarray}
\section{Evolution Equations}
The equations of motion for the particles are just the geodesic equations
(the characteristics of the \Vlasov\ equation).
These can be expressed in terms of the four momentum of 
the particle:
\begin{equation}
p^a \nabla _a p^b = 0,
\end{equation}
so in our coordinate system we have:
\begin{eqnarray}
\frac{dp^t}{d\tau}&=&-\Gamma^t{}_{tt}\,\left({p^t}\right)^2-2\Gamma^t{}_{tr}\,p^r p^t-
\Gamma^t{}_{rr}\, \left({p^r}\right)^2-
\Gamma^t{}_{\theta\theta}\,\frac{l^2}{r^4}\\
\frac{dp^r}{d\tau}&=&-\Gamma^r{}_{tt}\,\left({p^t}\right)^2-2\Gamma^r{}_{tr}\,p^r p^t-
\Gamma^r{}_{rr}\,\left({p^r}\right)^2-
\Gamma^r{}_{\theta\theta}\,\frac{l^2}{r^4}\\
\frac{dp^\theta}{d\tau}&=&-2 \Gamma^\theta{}_{r \theta}\, p^r p^\theta-
\Gamma^\theta{}_{\phi \phi}\, {p^\phi}
\end{eqnarray}
where the $\Gamma$'s are the Christoffel symbols defined by (\ref{Christoffel}), and 
$l^2={p_\theta}^2+{p_\phi}^2 / \sin^2{\theta}$ is the square of the magnitude of 
the angular momentum as before.
We want to consider the 4-momentum as a function of
our time coordinate and not the proper time of each particle. In order to do the 
transformation we can use 
the chain rule ($d/d\tau=
\left(dt/d\tau\right)\left(d/dt\right)$ and $p^t=dt/d\tau$):
\begin{eqnarray}
\frac{dp^t}{dt}&=&-\Gamma^t{}_{tt}\,p^t-2\,\Gamma^t{}_{tr}\,p^r -
\Gamma^t{}_{rr}\,\frac{\left({p^r}\right)^2}{p^t}-
\Gamma^t{}_{\theta\theta}\,\frac{l^2}{p^t r^4}\\
\label{dp^rdt1}
\frac{dp^r}{dt}&=&-\Gamma^r{}_{tt}\,p^t-2\,\Gamma^r{}_{tr}\,p^r -
\Gamma^r{}_{rr}\,\frac{\left({p^r}\right)^2}{p^t}-
\Gamma^r{}_{\theta\theta}\,\frac{l^2}{p^t r^4}\\
\frac{dp^\theta}{dt}&=&-2\, \Gamma^\theta{}_{r \theta}\, \frac{p^r p^\theta}{p^t}-
\Gamma^\theta{}_{\phi \phi}\, \frac{{p^\phi}^2}{p^t}
\end{eqnarray}
Furthermore it is convenient to express these equations in terms of $p_r$
rather than $p^r$:
\begin{equation}
p_r=g_{r\mu}p^{\mu}=a^2(t,r)\beta(t,r)p^t+a^2(t,r)p^r
\end{equation}
Solving for $p^r$:
\begin{equation}
\label{p^r}
p^r(t)=\frac{p_r(t)}{a^2(t,r)}-\beta(t,r)p^t(t)
\end{equation}
We can now compute the total derivatives with respect to coordinate time
as follows:
\begin{equation}
\frac{d}{d t}=\frac{\partial }{\partial t}+
                 \frac{d r}{d \tau}
                 \frac{d \tau}{d t}\frac{\partial }{\partial r}
             =\frac{\partial }{\partial t}+
                 \frac{p^r}{p^t}\frac{\partial }{\partial r}
\end{equation}
Applying this operator to equation (\ref{p^r}) we get:
\begin{equation}
\label{dp^rdt2}
\frac{dp^r}{dt}=\frac{1}{a^2}\frac{dp_r}{dt}
                 -2\frac{p^r}{a^3}\left(\frac{\partial a}{\partial t}
                 +\frac{p^r}{p^t}\frac{\partial a}{\partial r}\right)
                 -\beta \frac{dp^t}{dt} - p^t\left(
                 \frac{\partial \beta}{\partial t} + \frac{p^r}{p^t}
                 \frac{\partial \beta}{\partial r}\right)
\end{equation}
Substituting equations (\ref{p^r}) and (\ref{dp^rdt2}) into equation 
(\ref{dp^rdt1}) we obtain:
\begin{equation}
\frac{d p_r}{d t} =-\alpha \frac{\partial\alpha}{\partial r} p^t+
\frac{\partial \beta}{\partial r} p_r +
\frac{1}{a^3}\frac{\partial a}{\partial r}\frac{{p_r}^2}{p^t}+
\frac{l^2}{p^t r^3}
\end{equation}
which is the evolution equation for $p_r$. To derive the evolution equation
for $r$ we use the definition of $p^r$ ($p^r=dr/d\tau$) which after some
manipulation yields:
\begin{equation}
\frac{dr}{dt}=\frac{p_r}{a^2 p^t}-\beta
\end{equation}
To find the time component of the 4 momentum we make use of the 
normalization condition $p^\mu p_\mu=-m^2$:
\begin{equation}
\alpha p^t = \sqrt{m^2+\frac{{p_r}^2}{a^2}+\frac{l^2}{r^2}}
\end{equation}
It is also convenient, as previously mentioned, to use $\bar p^t=\alpha p^t$
rather than $p^t$ itself. Then the resulting geodesic equations in these coordinates are (we have
included equations for $p^\phi$, $p^\phi$ and $dp^\phi/dt$ because use we use 
them sometimes for visualization purposes):
\begin{eqnarray}
\label{evol1}
\frac{d p_r}{d t} &=&-\frac{\partial\alpha}{\partial r} \bar p^t+
\frac{\partial \beta}{\partial r} p_r +
\frac{\alpha}{a^3}\frac{\partial a}{\partial r}\frac{{p_r}^2}{\bar p^t}+
\frac{l^2 \alpha}{\bar p^t r^3}\\
\label{evol2}
\frac{d r}{d t}&=&\frac{\alpha p_r}{a^2 \bar p^t}-\beta\\
\label{evol3}
\bar p^t& =& \sqrt{m^2+\frac{{p_r}^2}{a^2}+\frac{l^2}{r^2}}\\
\frac{dp^\theta}{dt}&=&-\frac{2}{r}\left(\frac{\alpha p_r}{a^2 \bar p^t}
- \beta \right) p^\theta + \frac{1}{\tan{\theta}} \frac{\alpha}{\bar p^t}
\left(\frac{l^2}{r^4}-{p^\theta}^2\right)\\
\frac{d \theta}{dt}&=&\frac{\alpha p^\theta}{\bar p^t}\\
\label{evol4}
{p^\phi}^2&=&\frac{1}{\sin^2{\theta}}\left( \frac{l^2}{r^4} - {p^\theta}^2\right)\\
\label{evol5}
\frac{d \phi}{dt}&=&\frac{\alpha p^\phi}{\bar p^t}
\end{eqnarray}

\section{Conserved Quantities}
\label{conser}
Here we include the calculation of two quantities that we have used to check the code.
The first quantity that we discuss is a conserved quantity in 
flat spacetime related to the energy of the particles. This gives us a check of the evolution
when we fix the geometry to be Minkowskian (flat), 
which was useful in the development of the code.   
The second quantity that we calculate is a mass aspect function that coincides with the 
ADM mass at infinity. The conservation of the value of this function at the outer limit of our
range of integration gives another check in the case when we solve the fully coupled problem.

In flat space-time, the geometry is static, and therefore we have a time-like Killing
vector field. Choosing coordinates such that:
\begin{equation}
ds^2=-dt^2+ dr^2 + r^2\,d\theta^2+r^2\,{\mathrm sin}^2 \theta  d\phi^2, 
\end{equation}
we can write this Killing vector field as $\xi^\nu=\partial/ \partial t$.
If we contract $\xi_\nu$ with the stress energy tensor $T^{\mu \nu}$, we
get a vector field whose divergence is zero:
\begin{equation}
\nabla_\mu\left(T^{\mu \nu}\xi_\nu\right)=
\left(\nabla_\mu T^{\mu \nu}\right) \xi_\nu+T^{\mu \nu}\left( \nabla_\mu \xi_\nu \right)=0.
\end{equation}
The first term on the right hand side is zero by the conservation of the stress-energy
tensor, and the second is zero by the assumption that $\xi^\nu$ satisfies the Killing
equation\footnote{$\nabla_\mu \xi_\nu+\nabla_\nu \xi_\mu=0$}, and by the symmetry of 
$T^{\mu \nu}$.
Integrating this divergence over the four volume we get:
\begin{equation}
\int_V \nabla_\mu(T^{\mu \nu}\xi_\nu) \sqrt{-g}
d^4x = \int_{\partial V} T^{\mu \nu}
\xi_\nu n_\mu \sqrt{{}^{(3)} g}\ d^3x=0
\end{equation}
where the second integral is an integral over the 3-hypersurface $\partial V$ (boundary of $V$).
We choose this 3-hypersurface to be composed of two different hypersurfaces of constant $t$, $t=t_1$ and $t=t_2$,   
joined by a timelike-hypersurface at spatial infinity. We assume 
that the stress-energy tensor vanishes at spatial infinity, so that it does not contribute to the integral.
Let us also assume that $n_\mu=(-1,0,0,0)$ 
for the surface with $t_1$, and $n_\nu=(1,0,0,0)$ for $t_2$. 
In this case:
\begin{equation}
\int_{\partial V_{t_1}} T^{tt} \sqrt{{}^{(3)}g}d^3x = \int_{\partial V_{t_2}} T^{tt}
\sqrt{{}^{(3)}g}d^3x
\end{equation}
and since $t_1$ and $t_2$ are arbitrary this means that the integral is constant for any
value of the time and the integrand calculated using (\ref{tmunu}) 
\begin{equation}
\label{flat_conser}
C=\sum^N_{i=1} \frac{{p^t_i}^2}{m_i}
\end{equation}
is a conserved quantity \footnote{Not only (\ref{flat_conser}) is a conserved quantity but,
in flat spacetime, $p^t$ for each individual particle is conserved.}.

The other quantity that we calculate in this section is the mass aspect
function:
\begin{equation}
\label{mass}
M=\frac{r}{2}\left(1+\frac{\beta^2}{\alpha^2}-\frac{1}{a^2}\right)=
\frac{r}{2}\left(1+r^2K^\theta{}_\theta{}^2-\frac{1}{a^2}\right)
\end{equation}
To derive (\ref{mass}), we consider the transformation relating the metric (\ref{metric})
to the Schwarzschild metric (in a region of vacuum):
\begin{equation}
\label{smetric}
ds_s^2=-\left(1-\frac{2M}{r_s}\right)dt_s^2+\left(1-\frac{2M}{r_s}\right)^{-1}dr_s^2+r_s^2d\Omega^2
\end{equation}
Here the 's' subscript stands for Schwarzschild and we wish to
stress the point that the Schwarzschild time and our time coordinate are
different, while the radial coordinates are the same.
Following \cite{Shapiro:1985a} we get:
\begin{eqnarray}
dt_s&=&C\left(dt+Ddr\right)\\
dr_s&=&dr
\end{eqnarray}
where the last equation holds because both radial coordinates are areal.
Using these formulas in (\ref{smetric}) we get:
\begin{eqnarray}
\nonumber
ds_s^2=&&-\left(1-\frac{2M}{r}\right)C^2dt^2-2\left(1-\frac{2M}{r}\right)C^2
Ddtdr\\
\label{metric2}
&&-\left(1-\frac{2M}{r}\right)C^2D^2dr^2+\left(1-\frac{2M}{r}\right)^{-1}dr^2+r^2d\Omega^2
\end{eqnarray}
It is easy to see that in order for expressions (\ref{metric}) and
(\ref{metric2}) to be equal we need:
\begin{eqnarray}
C^2&=&-\frac{-\alpha^2+a^2\beta^2}{1-2M/r}\\
D&=&\frac{a^2\beta}{-\alpha^2+a^2\beta^2}
\end{eqnarray}
If we now compare the $dr^2$ terms of (\ref{metric}) and (\ref{metric2}),
we get an equation for the mass:
\begin{equation}
\frac{a^4\beta^2}{-\alpha^2+a^2\beta^2}+\frac{1}{1-2M/r}=a^2\, ,
\end{equation}
which after some manipulation can be rewritten as:
\begin{equation}
\frac{2M}{r}=1-\frac{1}{a^2}+\frac{\beta^2}{\alpha^2}\, .
\end{equation}
This last equation is interesting in its own, because when this
quantity, $2M(t,r)/r$, becomes equal to 1, we know that a marginally trapped surface
has been formed. We can see this by computing the expansion of the null geodesics
(see for instance \cite{Choptuik:phd}), which in this coordinates can be written as:
\begin{equation}
1-a(t,r)\,r\,K^\theta{}_\theta(t,r)\,=\,1-\frac{a(t,r)\,\beta(t,r)}{\alpha(t,r)}.
\end{equation} 
Therefore, if the expansion is zero, $1/a^2=\beta^2/\alpha^2$, and $2M(t,r)/r = 1$.

\resetcounters

\chapter{The Code}
\label{ch2}
In this chapter we explain the details of the method we have used 
to solve the equations which were presented in the last chapter.
\section{Particle-Mesh Methods}
We have used a method of calculation known as ``particle-mesh" (PM).
For a very complete review of this and other particle methods 
see \cite{Hockney:1981}.   

This technique solves the coupled Einstein-\Vlasov\ equations by splitting
 each time step into two stages: 
1) solution of the field equations on a mesh, and 2) updating of the 
positions of the particles via their equations of motion.


To provide some motivation for the use of PM methods let us first consider the
$N$ body problem in Newtonian gravity (where 
we assume $N$ is very large). We could calculate
the forces experienced by each particle $\vec F_j$ directly using 
\begin{equation}
\vec F_j=-G\sum_{i\ne j} \frac{m_i m_j}
{|\vec r_i-\vec r_j|^3}(\vec r_i-\vec r_j)
\end{equation}
Once the acceleration at each particle is known, we could integrate the
equations of motion over some small time interval $\Delta t$ giving the new positions 
and velocities of the particles. 
This method is denoted particle-particle, 
and computationally is very expensive since we have to perform $O(N^2)$
calculations per time step.

The ``particle-mesh" (PM) approach to the problem involves calculation of the mass density due
to the $N$ particles on a mesh (assuming each particle occupies a finite region 
in space). On this mesh, we can use finite difference approximation (or some 
other discretization technique, like a spectral method) to solve the Poisson 
equation for the gravitational potential:
\begin{equation}
\nabla^2 \Phi(t,\vec r) = 4\pi G \rho(t,\vec r).
\end{equation} 
Once the potential is found, we can interpolate the value of its derivative
to the {\em actual} particle positions to find the forces. A time step is 
then concluded by solving the evolution equations for each particle to provide updated 
positions and velocities.


PM methods are most effective when the close interactions between the particles are
not important for the evolution of the system. The main reason for this is that
smoothing of the particles
in a finite region of the space produce a large error in the density close to 
the particle positions. However, in cases when the dynamics is dominated by a 
{\it mean field} the  particle-mesh method is much faster than direct integrations 
like particle-particle.

In our case we want to solve the \Vlasov\ equation (equation (\ref{vlasov}))
by approximating the distribution function with a set of point particles. 
Further approximation is necessary to couple to the geometric equations,  
since a point particle gives rise to infinite densities.  
To solve the geometric equations we thus have to find the stress-energy tensor 
for the particles, where we consider each particle to be smoothed out over
a finite region of space. That is precisely the  PM method defined above. 

In the next sections of this chapter we will explain the specifics of 
the integration of the field equations, the evolution equations for each
particle and the interpolation scheme that we have used. We also include 
pseudo-code for the main routine of our program in Appendix D.

\section{The Field Equations}
We first explain how the equations (\ref{field1}-\ref{field4}) for the geometry
are solved numerically, assuming we know the quantities 
$\bar \rho$, $\bar \j_r{}$, $\bar S^r_r{}$, $\bar S^\theta_\theta{}$
\footnote{we will explain how to compute these quantities in section 3.4}.
The first two equations, equations 
(\ref{field1}-\ref{field2}), are
integrated from the origin, $r=0$, using the {\tt LSODA} \cite{lsoda}
integrator. The boundary conditions are given by the spherical symmetry of the
space-time, and by the demand that the spacetime be locally flat at $r=0$.
They are $a(t,0) = 1$, and $K^\theta{}_\theta(t,0)=0$. 

We compute the values of the functions $a_i$ and 
$K^\theta{}_\theta{}_i$ on a uniform grid of $N_r$ points at positions $r_i$:
$r_0=0$, $r_1=h$, $r_2=2h$,..., $r_{N_r}=r_{\mathrm max}$.  

In order to compute the values at $r=r_{i+1}$, we supply to {\tt LSODA} the values of the
functions at $r=r_i$ and the derivatives computed using equations (\ref{field1}-\ref{field2})
at $r=r_{i+1/2}$, using $\bar \rho$ and $\bar \j_r$ averaged between the i-th point and the
(i+1)-th point:
\begin{eqnarray}
\label{rhoaver}
\left[ \bar \rho \right] _{i+1/2} &=& \frac{1}{2} \left( \left[\bar \rho\right]_i + \left[\bar \rho\right]_{i+1} \right)\\
\label{javer}
\left[\bar \j_r\right]_{i+1/2} &=& \frac{1}{2} \left( \left[\bar \j_r\right]_i + 
\left[\bar \j_r\right]_{i+1} \right)
\end{eqnarray}
Once we have calculated $a$ we can solve the slicing equation:
\begin{equation}
\label{sli}
\alpha''=\alpha'(\frac{a'}{a}-\frac{2}{r})+\frac{2\alpha}{r^2}(a^2-1+2r\frac{a'}{a})+
4\pi a \alpha (\bar S^r_r+\bar S^a_a-3 \bar \rho),
\end{equation}
with the boundary conditions:
\begin{eqnarray}
\alpha^\prime(t,0)&=&0\\
\label{alphabcrmax}
\alpha(t,\infty)&=&1.
\end{eqnarray}
Here the first condition follows from (\ref{sli}) demanding that $\alpha$ be regular at $r=0$ 
and the second one follows from asymptotic
flatness, and the demand that $t$ measure proper time at infinity.
We use a second-order finite difference approximation on the previously defined mesh 
of points \{$r_i$\}:
\begin{eqnarray}
\frac{\alpha_{i+1}-2 \alpha_i+\alpha_{i-1}}{h^2}=&&\frac{\alpha_{i+1}-\alpha_{i-1}}{2 h}
\left(\frac{a_{i+1}-a_{i-1}}{2 h a_i} - \frac{2}{r_i}\right) +\\
 && \frac{2 \alpha_i}{{r_i}^2}
\left( {a_i}^2 - 1 +2 r_i \frac{a_{i+1}-a_{i-1}}{2 h a_i}\right)+\\
&&4 \pi a_i \alpha_i \left( \frac{\left[\bar S^r_r\right]_i}{{a_i}^2} +\left[\bar S^a_a\right]_i - 
3\left[\bar \rho \right]_i\right)
\end{eqnarray}
Rearranging this equation gives us:
\begin{equation}
\label{slieq}
\left( \frac{1}{h^2} + \frac{f_i}{2h} \right) \alpha_{i-1}-
\left( \frac{2}{h^2} + g_i \right) \alpha_i+
\left( \frac{1}{h^2} - \frac{f_i}{2h}\right) \alpha_{i-1}=0
\end{equation}
where:
\begin{eqnarray}
f_i&=& \frac{a_{i+1}-a_{i-1}}{2ha_i} - \frac{2}{r_i}\\
g_i&=& \frac{2}{{r_i}^2}\left( {a_i}^2-1+2 r_i \frac{a_{i+1}-a_{i-1}}{2 h a_i}\right)+
4 \pi a_i \left( \frac{\left[\bar S^r_r\right]_i}{{a_i}^2} +\left[\bar S^a_a\right]_i - 
3\left[\bar\rho\right]_i\right)
\end{eqnarray}
In addition to these equations we have the boundary equation at $r=0$:
\begin{equation}
\left(-3 + \frac{1/h^2 + f_2/(2 h)}{1/h^2 - f_2/(2 h)} \right) \alpha_1 +
\left(4 + \frac{-2/h^2-g_2}{1/h^2-f_2/(2h)} \right) \alpha_2 = 0
\end{equation}
which can be derived from the forward finite difference approximation to
$\alpha^\prime=0$ at $r=0$:
\begin{equation}
\frac{-3 \alpha_1 + 4 \alpha_2 - \alpha_3}{2h}=0\ ,
\end{equation}
and equation (\ref{slieq}) with $i=2$. We have also the boundary condition at 
$r=r_{\mathrm max}$:
\begin{equation}
\label{l_rmax}
\alpha_{N_r} = \left[\sqrt{1 - \left(\frac{2 M}{r}\right)}\right]_{N_r}
\end{equation}
This is an approximation to (\ref{alphabcrmax}) at a finite value of $r=r_{\mathrm max}$,
where $M$ is the mass aspect function defined by equation (\ref{mass}).
These equations form a linear system of algebraic equations that can be solved using a 
tridiagonal solver (we have used the LAPACK \cite{LAPACK} routine {\tt dgtsv}).
\\
\section{The Evolution Equations}
\label{evolution}
To evolve the particles' positions and momenta we integrate the geodesic equations 
(\ref{evol1}-\ref{evol2}). The values of the coefficients in these equations 
(basically products and quotients of $a$, $\alpha$, $\beta$, $a^\prime$, $\alpha^\prime$
 and $\beta^\prime$)  
must be calculated at the particles' positions using the values obtained at the mesh points 
\{$r_i$\}. The mesh values are interpolated to the particle positions using the same operator kerner
used to produce mesh values from particle quantities (this procedure is explained 
in section \ref{interpolation}). These equations are integrated using the {\tt LSODA}
integrator as in the case of the Hamiltonian and momentum constraints 
(\ref{field1}-\ref{field2}). For a given particle at position $r^n$ and momentum $p_r^n$ at
time step $n$, we calculate the new position $r^{n+1}$ and momentum $p_r^{n+1}$ at $t=t^{n+1}$ by
suplying to {\tt LSODA} the values 
of $r^n$, $p^n$ and their derivatives at time step $t=t^n$. 
This procedure has an accuracy $O(\Delta t)$ (since the derivatives are calculated 
using the metric coeficients at $t=t^n$) where
$\Delta t=t^{n+1}-t^n$.  
In our program we chose a value of $\Delta t$ proportional to
$h$, i.e. $\Delta t = \lambda\, h$, where usually $\lambda = 1.0$.


We need to take special care if a particle leaves the range of integration 
($r>r_{\mathrm max}$) or if it reaches the origin. In the first case we stop 
considering the particle and
we change the total number of particles to $(N-1)$. 
When a particle $p$ reaches the origin, in other words when $r_p<0$, we reflect 
the particle, i.e. we set:
\begin{eqnarray}
\left[r \right]_p &\Longrightarrow& -\left[ r \right]_p \\
\left[p_r \right]_p &\Longrightarrow& -\left[ p_r \right]_p \\
\left[l \right]_p &\Longrightarrow& \left[ l \right]_p
\end{eqnarray} 
As mentioned previously, equations (\ref{evol4}), for $p^\phi$, and (\ref{evol5}), for $\phi$, 
are also sometimes solved for visualization purposes and is also integrated in the same way.   
\section{Interpolation and Restriction}
\label{interpolation}
In this section we explain how we calculate the energy distributions on the mesh
from a given set of particles. In other words we will discuss the different
possibilities for the interpolation operator $W(r-r_i)$ appearing in equation 
(\ref{inter}). At the end of the section we explain how to use the same  
operators to restrict the coefficients of the geodesic equations calculated
on the mesh (i.e. the geometric quantities), to the particles' positions.

We can define a hierarchy of interpolation operators which can be classified by
how many derivatives of the resulting density (interpolant) are continuous. 
The lowest-order interpolation (the resulting density is discontinuous, piecewise constant), 
called nearest grid point interpolation, 
assigns the density of each particle to its nearest grid point (NGP). 
Thus the interpolation kernel can be written as: 
\begin{equation}
W(x-x_p)= \left\{ \begin{array}
{r@{\quad:\quad}l}
1 & |x-x_p| \le h/2 \\ 0 & {\mathrm {otherwise}} \end{array} \right.
\end{equation}
\begin{figure}[htbp]
\epsfxsize=4.0in
\centerline{\epsfbox[35 180 535 420]{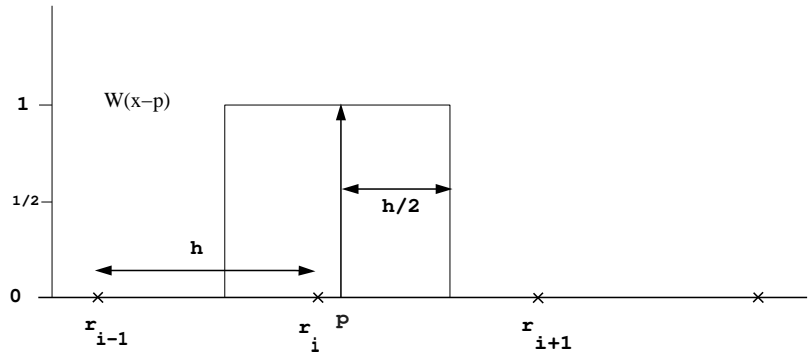}}
\caption[Kernel for the first order interpolation (NGP).]
{Kernel for the first order interpolation (NGP).}
\label{ngp}
\end{figure}
where $x_p$ is the position of the particle and $h$ is the mesh spacing 
of the grid on which we solve the field equations. 

Figure \ref{ngp} displays a typical NGP
kernel operator for a particle at position $p$, and
a grid with grid points at $r=r_{i-1}$, $r_i$, $r_{i+1}$, etc. This kernel will produce a
density distribution given by:
\begin{equation}
\rho(x) = \sum_{i=1}^{N} \rho_p \  W(x-x_p) 
\end{equation}

A commonly used higher order interpolation is called ``cloud in cell" 
(CIC) interpolation 
which distributes the density of each particle into two grid cells. 
In this case the resulting density is piecewise linear and the kernel takes the
form: 
\begin{equation}
W(x-x_p)= \left\{ \begin{array}
{r@{\quad:\quad}l}
1-|x-x_p|/h & |x-x_p| \le h \\ 0 & {\mathrm {otherwise}} \end{array} \right.
\end{equation}
\begin{figure}[htbp]
\epsfxsize=4.0in
\centerline{\epsfbox[35 180 535 410]{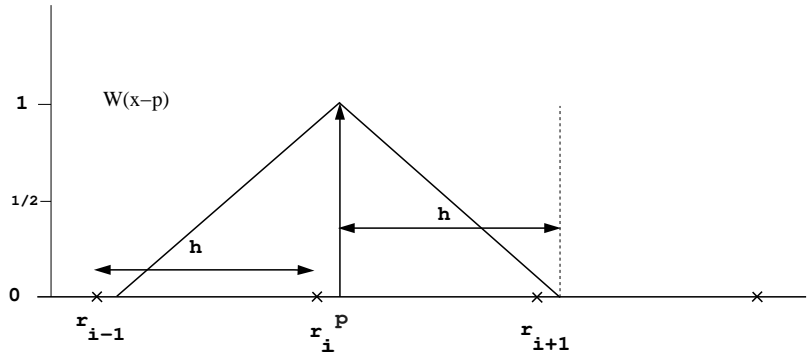}}
\caption[Kernel for the second order interpolation (CIC).]
{Kernel for the second order interpolation (CIC).}
\label{cic}
\end{figure}
Figure \ref{cic} is a graph of the kernel of the CIC interpolation. 
In our calculations we have used the CIC interpolation operator.

To restrict the Christoffel symbols (and other geometric quantities)
calculated on the mesh to the particles' positions we use the same
kernel in the following way:
\begin{equation}
\label{restr}
F(x_p)=\sum^{N_r}_{i=1} F(x_i) W(x_i-x_p)
\end{equation}
where $x_p$ is the position of the particle, $x_i$ are 
the grid points, and $F$ is any of  the coefficients that appear in the evolution equations. 
The coefficients $F$ are products and quotients of the metric coefficients and their derivatives.
In order to calculate derivatives we use the following finite difference approximation on the mesh:
\begin{equation} 
\left[q^\prime\right]_i=\left( q_{i+1}-q_{i-1} \right) / h + O(h^2)
\end{equation}
and then use equation (\ref{restr}) to find an approximate value of $q^\prime$ 
at the particle's position.

\section{Initial Set of particles.}
To initialize the set of particles which we evolve, we specify the
particle distribution (number of particles per unit of areal coordinate) 
and the velocity distribution (specifically the number of particles per $p_r$ and $l$).
This correspond to a distribution function in the sense of (\ref{distri}) 
which is separable in the $r$, $p_r$ and $l$ variables. In other words:
\begin{equation}
\label{distris}
f(r,p_r,l)=R(r)\,P(p_r)\,L(l)
\end{equation}

Moreover, instead of specifying $P$ as a function of $p_r$ we specify $P=P(\bar p_r)$ where 
$\bar p_r=p_r/a$. This allow us 
to calculate the value of $p^t=\sqrt{m^2+\bar p_r^2 +l^2/r^2}$, and therefore $\bar \rho$,
$\bar \j_r$ and $\bar S$, for each particle without knowing the geometry in advance.  
This allows us to decouple the tasks of specify initial data for the particles, and ensuring
that the initial data satisfies the constraint equations.

We use 1-dimensional Monte Carlo techniques applied to each of the functions $R(r)$, $P(p_r)$,
 $L(l)$ to get an specific ensemble of the particles.
In theory  the statistical error that we produce goes like 
$O(N^{-1/2})$, where $N$ is the number of particles
that we use to sample the distribution function (see section \ref{errors}).

\resetcounters

\chapter{Checks}
\label{ch3}
In this chapter we explain different checks we have performed to test that we have properly
implemented the algorithm.
The first of these test was checking if the 
quantity given by equation (\ref{flat_conser}) was conserved in the flat spacetime case.  
In our code this value was conserved up to the tolerance of the {\tt LSODA} integrator
for initial data which was gaussian in both the radial and velocity distributions. We have also checked that 
a test particle moving in a fixed background follows the geodesic of the background, and that its energy is conserved.

The rest of the checks that we describe here are for the fully 
coupled problem. In section \ref{mass_conser} we discuss the conservation of 
the mass aspect function at the outer boundary ($r=r_{\mathrm max}$).
We also have simulated certain clusters of particles in circular motion known as
Einstein clusters;
results from these simulations are discussed in \ref{einstein_clusters} 
in section \ref{einstein_clusters}. 
At the end of the chapter we study how the numerical errors scale with the number 
of particles we use to sample the distribution function $f$,
and as well as with the mesh spacing $h$.
\section{Mass Conservation}
\label{mass_conser}
\begin{figure}[htbp]
\epsfxsize=3.0in
\centerline{\epsfbox[100 180 600 700]{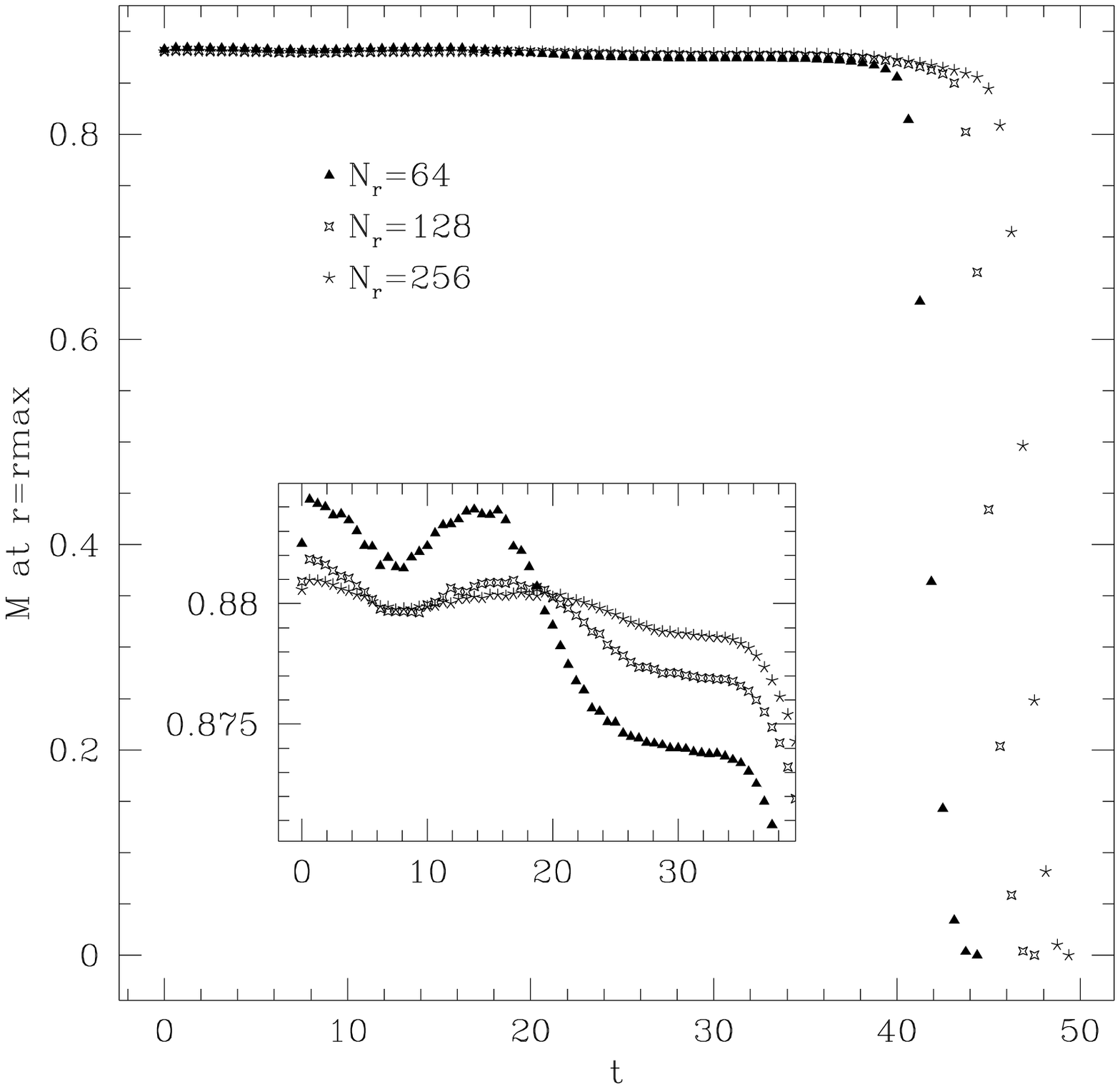}}
\caption[Value of $2M(r,t)/r$ at $r=r_{\mathrm max}$ as a function of time.]
{Value of $2M(t,r)/r$ at $r=r_{\mathrm max}$ as a function of time.}
\label{mass_conservation}
\end{figure}

The value of the mass function given by equation (\ref{mass}) at 
infinity should be a conserved quantity if the spacetime is asymptotically
flat (see for instance \cite{MTW}).
As an approximation to the value at infinity we can monitor 
the value at the maximum value of the radial coordinate, $M(t,r_{\mathrm max})$.
 This
check will make sense as long as no matter leaves the 
range of integration (note that energy cannot escape in the form of radiation since that 
is forbidden in spherical symmetry). In Figure \ref{mass_conservation}
we plot $M(t,r_{\mathrm max})$ 
for three different mesh scales ($N_r=64$, $N_r=128$ and
$N_r=256$) and with a fixed number of particles, $N=10 000$.  

We observe that the value of $M$ is conserved until $t \approx 40$ which s the time 
when the particles begin to leave the range of integration. In the inset we see a detail
of the function between $t=0$ and $t=40$ showing that the conservation 
is better than $1 \%$ for $N_r = 64 $ and improves somewhat as we increase the 
number of grid points. Theoretically (and in the limit $N \to \infty$) we should observe 
that the variation of the value of mass 
should decrease linearly with the number of points, i.e. first order. Although the convergence test 
here is not as definitive as it tends to be for pure finite-difference methods, the inset of 
Fig.~\ref{mass_conservation} provides fairly good evidence that our code's mass 
conservation {\em would} converge linearly with $h$ in the continuum limit.

\section{Einstein Clusters}
\label{einstein_clusters}
Probably the strongest check of the reliability of our code is 
its ability to simulate clusters of particles that only have rotational motion 
(zero radial component of the 4-momentum). Let us calculate what 
initial conditions we must set in order to obtain such a cluster.
 
\begin{figure}[htbp]
\epsfxsize=3.0in
\centerline{\epsfbox[100 150 600 700]{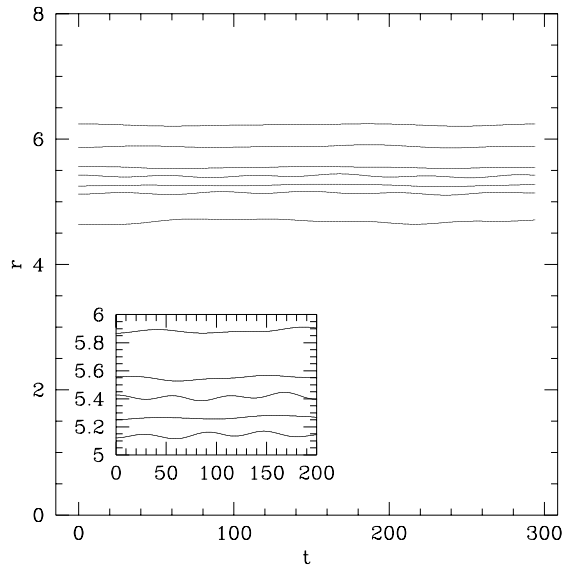}}
\caption[Radial coordinate of 7 particles as a function of time for a cluster
with gaussian energy density profile.]
{Radial coordinates of 7 particles as a function of time for a static cluster with gaussian
energy density profile.}
\label{trayectories}
\end{figure}

Assuming we know what the energy density of the cluster of particles is,
let us calculate the velocity profiles  required in order to produce a cluster in 
equilibrium. In the process of obtaining the velocity profiles, we will also 
find the geometry of the spacetime created by the cluster. 
The metric coefficient $a(r)$ can be calculated using equation (\ref{mass}). If the
spacetime is static then $\beta(r)=0$ and the resulting equation is:  
\begin{equation}
a(r) =\left( 1-2\frac{M(r)}{r} \right)^{-1},
\end{equation}  
where we calculate $M(r)$ by integrating the density $\rho(r)$ over the radial coordinate,
\newline
$M(r)=\int 4 \pi r^2 \rho(r) dr$.
To calculate the lapse function we can make use of the evolution equation 
for the extrinsic curvature (\ref{evolK}), which in this case reduces to:
\begin{equation}
\left(\frac{\alpha r}{a}\right)^\prime=\alpha a
- 4 \pi \alpha a r^2 \rho.
\end{equation}
After a little bit of manipulation this equation can be written as:
\begin{equation}
\label{stat_alpha}
\frac{\alpha^\prime}{\alpha}=-\frac{1-a^2}{2r}=
\frac{M(r)}{r^2}\frac{1}{\left(1-2M(r)/r\right)}.
\end{equation}
We want to find the angular momentum $l(r)$ needed to get the particles
rotating with zero radial component of the 4 momentum. Imposing that 
condition ($p_r=0$ and $\partial p_r /\partial t = 0$) on the geodesic equation 
for $p_r$, equation (\ref{evol1}), we obtain:
\begin{equation}
-\frac{\partial \alpha}{\partial r} \bar p^t + \frac{l^2 \alpha}{r^3 \bar p^t}=0
\end{equation}
Taking into account that  
\begin{equation}
\bar p^t = \sqrt{m^2 + \frac{l^2}{r^2}}\ ,
\end{equation}
and using equation (\ref{stat_alpha}), we find the equation for $l(r)$:
\begin{equation}
\label{angularm}
l^2(r) = m^2 r^2 \left(\frac{M(r)}{r}\right)
\frac{1}{\left(1-3 M(r)/r\right)}
\end{equation}
This is the same expression as the angular momentum for a particle in
circular orbit around a Schwarzschild black hole with mass $M(r)$ (see \cite{Wald}).    
In Figure \ref{trayectories} we have plotted the areal coordinate for 
some particles in a cluster with gaussian energy density profile and total mass 
$M \approx 0.57$ as a function of time. Further we can see that the particles retain 
roughly the same radial coordinate for a time longer than $500 M$ (2 complete orbits).  
Also, the motion appears to be stable since the particles oscillate radially with roughly constant amplitude 
(see inset).
We found that for clusters with gaussian energy densities and initial maximum $6 M/r = 1.01$, 
the cluster becomes unstable and formed
a black hole, whereas for clusters with initial maximum $6 M/r= 0.97$ the cluster was 
stable\footnote{This cluster actually achieves  gets $6 M/r > 1$ during certain parts of 
the evolution.}. This result
agrees with the result that Einstein clusters become unstable for $6M/r=1$ (see
\cite{Shapiro:1985a} and references therein).  

We have also simulated distorted static clusters for which we specify $(1-\epsilon)$ 
times the angular momentum fixed by (\ref{angularm}); this allows us to control 
the deviation from the {\em ideal} static case. We have observed results similar  
to the case just described where the perturbation is effectively introduced by the numerical approximation 
itself;
i.e. our distorted clusters tend to be stable when 
\begin{equation}
\max_{t,r} \left(6 M(t,r)/r \right) < 1.
\end{equation}
\section{Numerical Errors.} 
\label{errors}
There are two principal sources of numerical errors in our code:
1) sampling of the initial distribution function
with a finite number of particles (statistical error), and 
2) approximation of the derivatives in the field equations 
by finite-difference operators
(truncation error).

In the continuum limit --- where the number of particles, $N$, approaches infinity, 
and the mesh spacing, $h$, goes to $0$ --- 
we need to check that these numerical errors
go to zero to establish that, in this limit, we {\em are} solving the continuum 
system of equations (\Vlasov\ equation coupled with Einstein's equations). 

First let us study how the statistical error varies with the number of particles. 
We can estimate the statistical error in one function by subtracting  
the functions computed using two different initial sets of particles
(two different sets, each sampled using a Monte Carlo method and the same 
distribution function).
Here, the specific function which we monitor is $2M(t,r)/r$, and we take the $L_2$ norm of
the difference to be an estimate of the statistical error at each time.
In other words, if 
${}^{(1)}\left(2M(t,r)/r \right)$ and
${}^{(2)}\left(2M(t,r)/r\right)$ are calculated from two different sets of particles
(generated from the same distribution function), then
the statistical error estimate at a given time $t_o$ is computed by:
\begin{equation}
\Delta s(t_o,r) = {}^{(1)}\left(2M(t_o,r)/r\right) - {}^{(2)}\left(2M(t_o,r)/r\right).
\end{equation}  
Furthermore, we take the $L_2$ norm of this function:
\begin{equation}
\Delta s(t_o)= \sqrt{\sum_{i=1}^{N_r} \Delta s(t_o,r_i)^2 / N_r}
\end{equation}  
\begin{figure}[htbp]
\epsfxsize=3.0in
\centerline{\epsfbox[80 200 600 700]{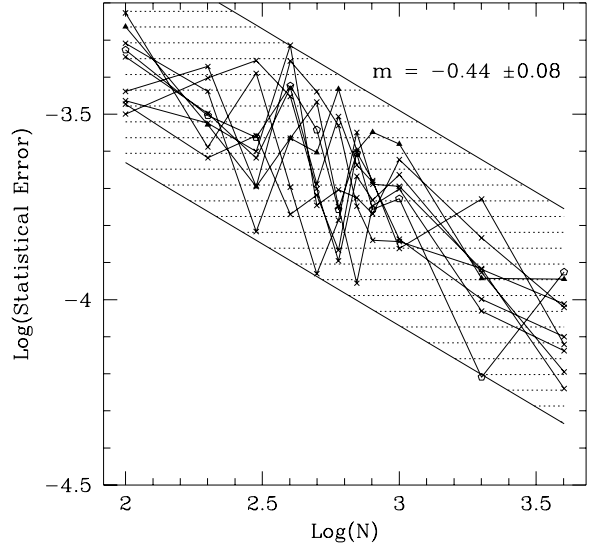}}
\caption[Statistical Error.]
{Statistical Error.}
\label{mc_err}
\end{figure}
In Figure \ref{mc_err} we have plotted these estimates as a function of the 
number of particles at the initial time $t_o=0$ and for fixed mesh spacing, h. 
We have displayed 9 different estimates for the statistical error, and the shaded area of the graph
shows bounds for the ensemble.
The average slope is $m=-0.44 \pm 0.08$ where the
uncertainty is the standard deviation calculated from the individual slopes. 
We see that these results roughly agree with the expectation that the statistical 
error should scale as $N^{-1/2}$. 

\begin{figure}[htbp]
\epsfxsize=3.0in
\centerline{\epsfbox[80 200 600 700]{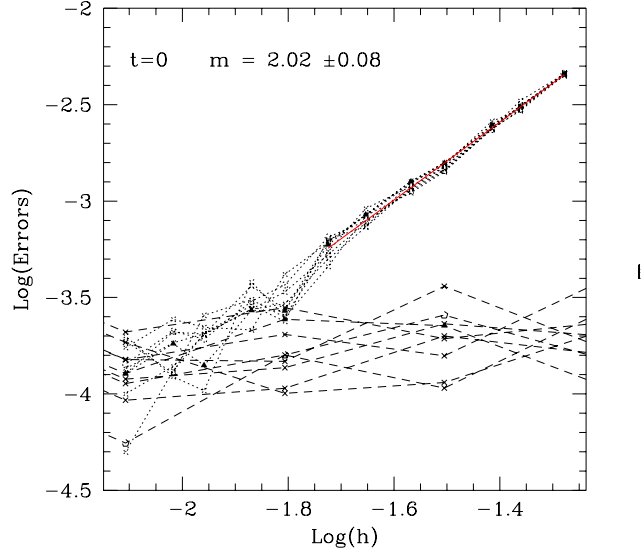}}
\caption[Truncation (dotted lines) and statistical error estimates (dashed lines) at t=0.]
{Truncation (dotted lines) and statistical error estimates (dashed lines) at t=0.}
\label{te_t0}
\end{figure}

Let us turn our attention now to the truncation error.
We assume a Richardson expansion for a mesh function $q^h(t,r)$, calculated with
a given mesh spacing $h$:
\begin{equation}
q^h(t,r)=q(t,r)+\Delta q_s(t,r)+h^p e_p(t,r)+O(h^{p+1})
\end{equation}
where $\Delta q_s$ is the statistical error. We estimate the truncation error in $q^h(t,r)$ 
via
\begin{equation}
q^h(t,r)-q^{h^\prime}(t,r)
\end{equation}
where $h^\prime < h$, specifically we have used log$(h^\prime)\approx-2.4$).
In this way we use the function calculated with $h^\prime$
as a better approximation (a reference) to the analytic solution than the one with larger $h$. 
It is important that the statistical error $\Delta q_s \ll h^p e_p$
otherwise this recipe to calculate the estimate
for the truncation error is meaningless since $q^h(t,r)-q^{h^\prime}(t,r)$ 
would be dominated by the statistical error.  

In Figure \ref{te_t0} we show the base ten logarithm  of the $L_2$ norm of the 
estimates of both the truncation and the statistical errors at $t=0$. 
We have plotted the truncation error estimates with dotted lines, 
and the statistical ones with dashed lines. 

In these calculations
we have not fixed the total number of particles, but rather the number of particles per grid cell
(the value we use in this particular case is 500 particles per grid cell on average). 
The reason for fixing the number of 
particles per cell, rather than the total number of particles, is that the statistical
 error scales with $h$ (for the total number 
of particles fixed) roughly as $h^{-1}$. Increasing the 
number of particles as we decrease $h$, we compensate for this bias by 
maintaining roughly constant and small statistical error.

In the part of the Figure where the statistical error is much smaller than the truncation error
we have calculated the average slope for the truncation error (the solid line in the Figure).
The resulting slope is $m=2.02 \pm 0.08$ where
again the uncertainty is the standard deviation. At $t=0$ we thus observe 
second order convergence which is in
agrement with the scheme that we have used to compute the geometric quantities.

\begin{figure}[htbp]
\epsfxsize=3.0in
\centerline{\epsfbox[80 200 600 700]{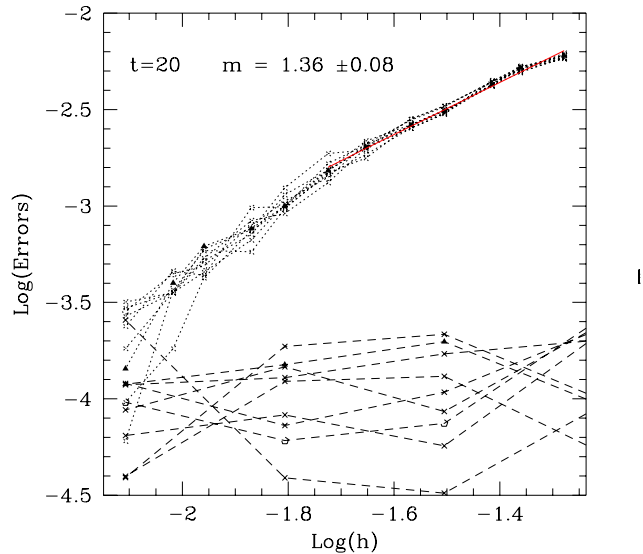}}
\caption[Truncation (dotted lines) and statistical error estimates (dashed lines) at t=20.]
{Truncation (dotted lines) and statistical error estimates (dashed lines) at t=20.}
\label{te_t20}
\end{figure}
Figure \ref{te_t20} shows the results of our error analysis at $t=20$. 
We see that in this case the slope
is not $2$ but smaller, $m=1.36 \pm 0.08$. This is consistent with the observation made in section \ref{evolution} that the numerical scheme that we have used
for the evolution is only first order accurate in $h$.


\resetcounters

\chapter{Results}
\label{ch4}
In this chapter we summarize the main results we have obtained.
In order to find the critical solutions (solutions sitting at the threshold of black hole
 formation) we have performed 
bisection searches using the total rest mass (sum of the rest masses of the individual particles)
of the cluster, $M_o$, as the search parameter $p$.
In section \ref{static_solution} we discuss some characteristics of the critical solutions.
In 5.2 we give some indications for the existence of a  scaling 
laws, $\tau \sim -\sigma \ln{|p-p^*|}$, and we discuss the possibility of universality
in the model in section \ref{universal}. The chapter ends with some conclusions and some ideas for future work.

\section{Is the critical solution Static ?}
\label{static_solution}

In this section we present evidence that, for initial data with non-zero angular momentum,
as we approach criticality 
the solution for the space time approaches a static solution. 

\begin{figure}[htbp]
\epsfxsize=3.5in
\centerline{\epsfbox[20 150 525 650]{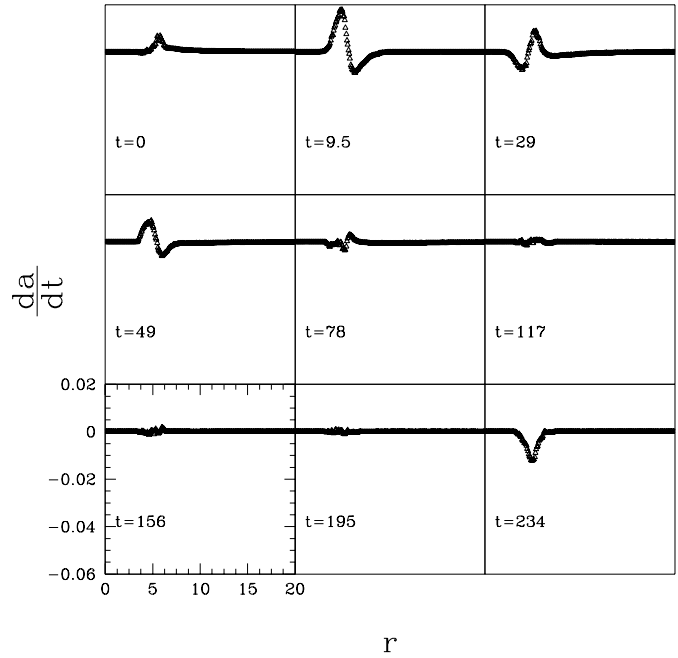}}
\caption[Snapshots of $da/dt$.]
{Snapshots of $da/dt$.}
\label{dadt1}
\end{figure}

  
We will first focus on one family of initial conditions and afterwards we will 
try to explain what we observed for different families.
The initial distributions on equation (\ref{distris}) are:
\begin{eqnarray}
\label{gaussian1}
R(r)&\propto&r^2 e^{\left(-(r-r_o)^2/\Delta_r^2 \right)}\,\Theta(r)\\
P(p_r)&\propto&e^{\left(-(\bar p_r-{\bar p_r}{}_o)^2/
                       {\Delta_{\bar p_r}}^2 \right)}\\
\label{gaussian2}
L(l)&\propto&e^{\left(-(l-l_o)^2/{\Delta_l}^2 \right)}\,\Theta(l)
\end{eqnarray}
where $\bar p_r=p_r/a$, and $\Theta$ is the step or Heaviside function. 
For $r_o=5$, $\Delta_r=1$, $\bar {p_r}_o=0$, $\Delta_{\bar p_r}=2$,
$l_o=12$, $\Delta_l=2$ (we will call this family ``almost" time symmetric
because the gaussian for the radial momentum is centered around $p_r=0$)
we found the critical solution for a cluster rest mass 
about $M_o \approx 1.3$.  
In Figure \ref{dadt1} we show a few snapshots 
from the evolution of $da/dt$ resulting from initial data which
 is close to criticality but which eventually disperses. 
At early times, $da/dt$ oscillates, but between $t=117$ to $t=195$ 
it seems to get close to zero. 
In the last snapshot ($t=234$) we observe how $da/dt$ has become negative, 
corresponding to dispersal of the particles.
We observe similar behavior for the time derivatives of 
all the metric coefficients.

\begin{figure}[htbp]
\epsfxsize=3.5in
\centerline{\epsfbox[20 185 550 675]{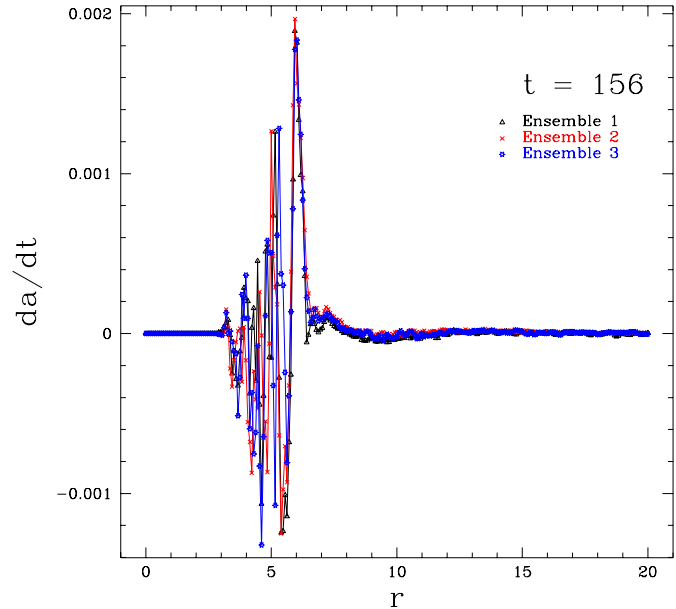}}
\caption[Three ensembles of $da/dt$ at $t=156$.]
{Three ensembles of $da/dt$ at $t=156$.}
\label{dadt_156}
\end{figure}

\begin{figure}[htbp]
\label{maxtmr}
\epsfxsize=3.5in
\centerline{\epsfbox[50 190 550 675]{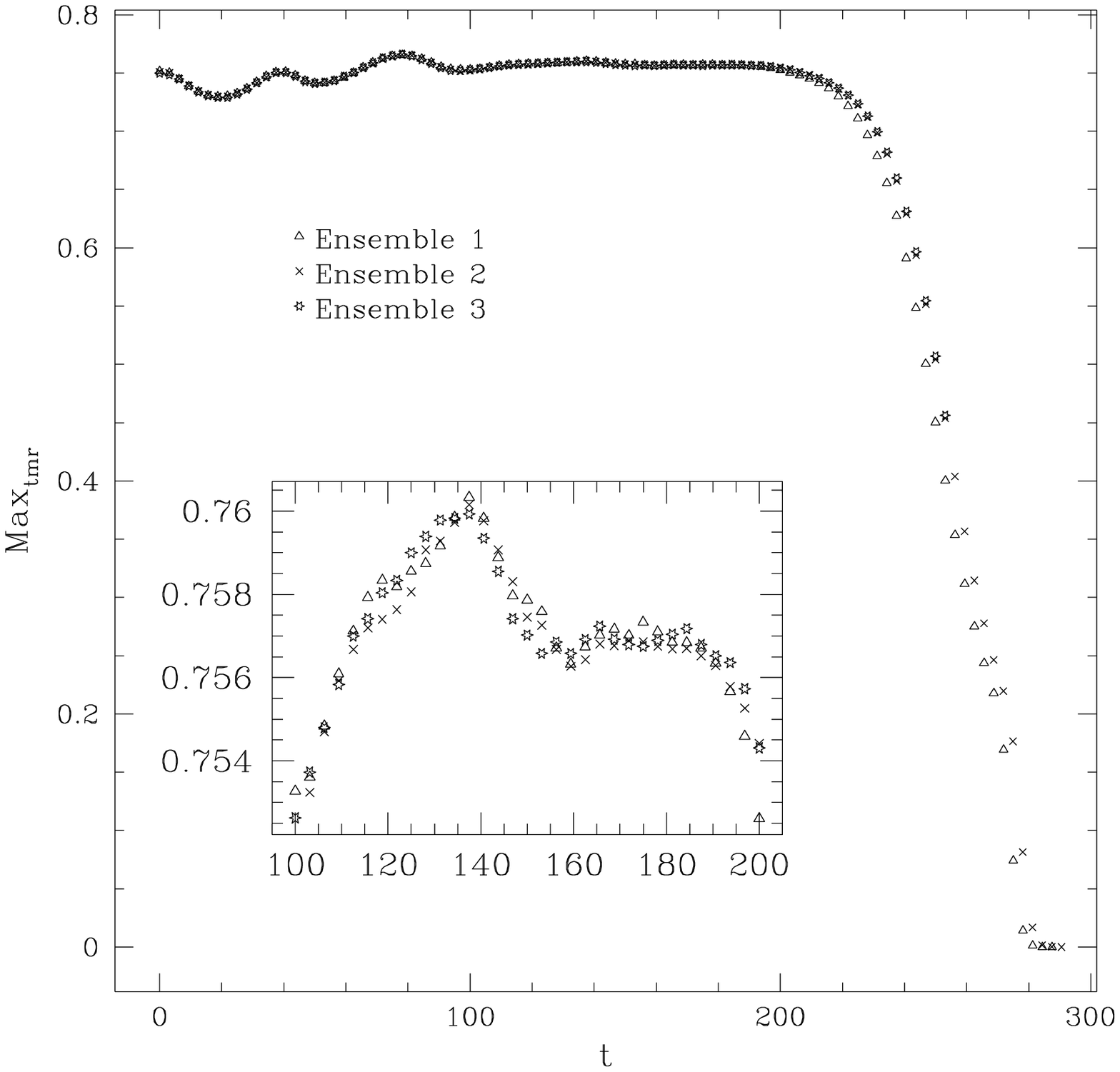}}
\caption[$\max_r \left( 2M(t,r)/r \right)$ as a function of time.]
{$\max_r \left( 2M(t,r)/r \right)$ as a function of time.}
\label{dadt}
\end{figure}

In order to see how close to zero $da/dt$ becomes, we show in Figure
\ref{dadt_156}, a detail of $\dot a(t,r)$  at 
$t=156$ for three different sets of particles. The 
statistical error is roughly of the same order as the function itself
(this is not true from $r\approx6$ to $r\approx8 $ where the three ensembles
agree on a non-zero value for the derivative, however we suspect that this feature 
will decrease if we tune closer to the critical solution and if we use greater resolution). 
This shows
that the derivative of the function could be zero as we anticipate
but it doesn't give us a definite answer. The value of 
$da/dt$  could be smaller that the statistical error.  
In order to definitively answer this question we will need to have better 
resolution and
decreased statistical errors.
As we have commented in section \ref{conser} an interesting
function we monitor is $2M(t,r)/r$. This function is
a scale invariant
quantity which tells us a how close we are to trapped surface formation. 
We plot  
\begin{equation}
\max_r \frac{2 M(t,r)}{r}
\end{equation}
in Figure \ref{maxtmr}. 
In the inset we show a detail for the period of time when the
maximum seems to have a roughly constant value. Again, between $t=140$ 
and $t=190$ the statistical errors are of the same order as the fluctuations
in $2M(t,r)/r$

If we accept that the metric coefficients are independent of our time
coordinate then we have a stationary space-time. If, in addition, the  
vector $N^a=d/dt$ introduced in Chapter (\ref{ch1}) is orthogonal to the 
spatial hypersurfaces, then the space-time is static. This
then  implies that $\Omega_a=N_a$, or that the shift vector, and in particular the 
shift function $\beta(t,r)$, is zero.

\begin{figure}[htbp]
\epsfxsize=3.5in
\centerline{\epsfbox[20 150 525 650]{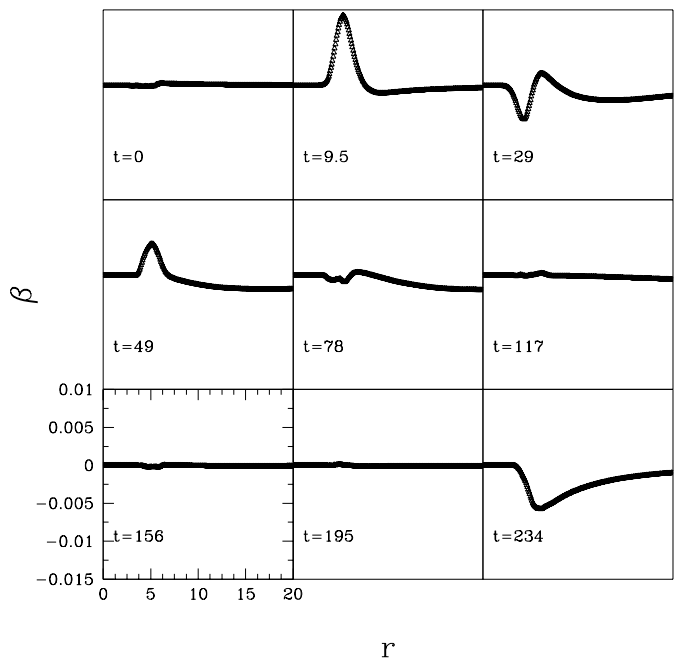}}
\caption[Shift function as a function of time.]
{Shift function as a function of time.}
\label{b}
\end{figure}

\begin{figure}[htbp]
\epsfxsize=3.5in
\centerline{\epsfbox[20 150 525 650]{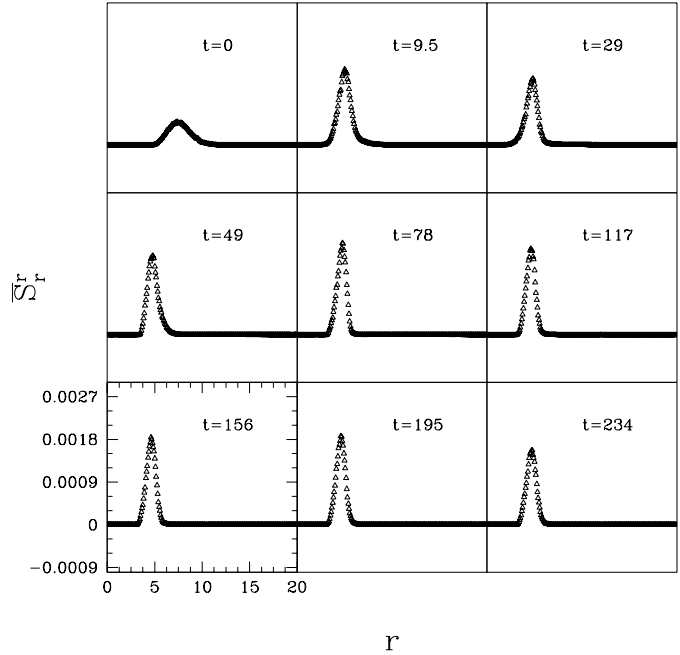}}
\caption[Evolution of $\bar S_r(t,r)$.]{Evolution of $\bar S_r(t,r)$.}
\label{Sr}
\end{figure}

In Figure \ref{b} we show the evolution of the shift function. 
During the period when the time derivatives of the metric
coefficients are close to zero, the shift function $\beta(t,r)$ also is close to 
zero, or at least of the same order as the statistical fluctuations as before.
Thus the solution for the space-time sitting at the threshold of  
black hole formation could be a static solution (if it is not static, the amplitude
of the derivatives and the shift function have small amplitudes relative to the statistical
error).
 We also note that the total current density $j_r$ also tends to be of the order of 
the statistical fluctuations, but that the $S^r_r$ component of the stress energy 
tensor is not zero in the critical regime, as we can see in Figure
\ref{Sr}. Therefore, on average there are the same 
number of particles with positive (outwards) and negative (inwards) $r$ 
components of the 4-momentum but  the absolute value of $p_r$ is not
zero on average. 

This result and the fact that the maximum value of $2M(r)/r$ for this critical solution
is $\sim 0.76$ indicate that this {\em cannot} be one of the Einstein clusters we have considered
previously since there are no equilibrium clusters (either stable or unstable)
with maximum $2M(r)/r$ larger than $2/3$.

\newpage

\section{Time Scaling}
A Type I critical solution usually approaches a static or a periodic solution.
As was argued in the previous section, we believe the critical solution in the current case 
is static.  
As we approach $p^*$ the dynamical solution spends more and
more time close to this putative static solution.

\begin{figure}[htbp]
\epsfxsize=3.2in
\centerline{\epsfbox[50 165 560 690]{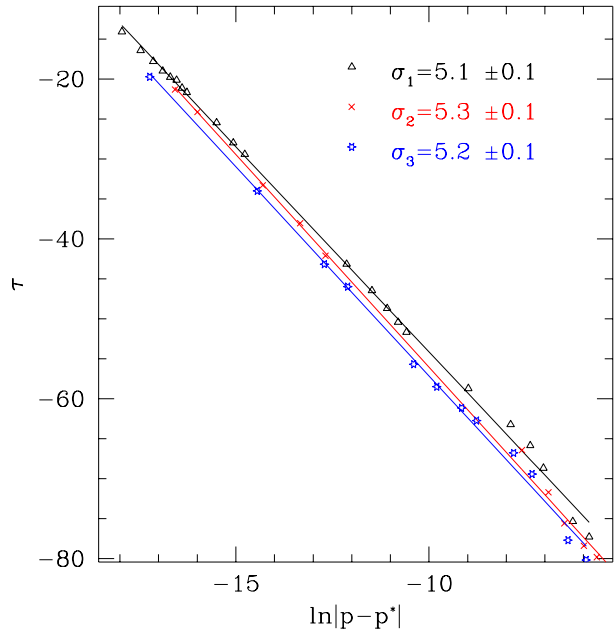}}
\caption[Scaling law for the time. The error for each value of $\sigma$ is the standard deviation of the slope computed using a least squares fit.]{Scaling law for the time. The error for each value of $\sigma$ is the standard deviation of the slope computed using a least squares fit.}
\label{scaling}
\end{figure}

We have calculated the time $t$ that each of the dynamical solutions spends
inside some radius $r=r_o$ (in particular $r_o=6$).
Figure \ref{scaling} shows the
plot of $\tau=t-t_c$ versus ln$(p)$ (natural logarithm)
where $t$ is the time that the solution with search parameter $p$ spends 
inside $r=r_o$, and $t_c$ is the same quantity for 
the solution closest to criticallity. 

We show three different ensembles of particles for
the gaussian family given by equations (\ref{gaussian1}-\ref{gaussian2}) 
and again using ``almost time symmetric" initial data.
We also have normalized the critical solution in such a way that 
it has ADM mass equal to $1$ at 
infinity\footnote{This will be
useful when we compare with other families of initial data.}, i.e.:
\begin{eqnarray}     
\label{sca1}
r &\rightarrow& r /M^c(t^*,r_{\mathrm max})\\ 
\label{sca2}
t &\rightarrow& t /M^c(t^*,r_{\mathrm max}),
\end{eqnarray}
where $M^c(t^*,r_{\mathrm max})$ is the value of the mass aspect function
at $r=r_{\mathrm max}$ for the solution closest to criticality during the critical regime.
We can see that there is a {\it rough} linear relation between the 
time $\tau$ and ln$|p-p^*|$ with slope $-(5.2 \pm 0.2)$ (the error is an estimate of
the statistical error between different sets of particles):
\begin{equation}
\label{scal_eqn}
\tau \sim  -(5.2 \pm 0.2) \ln|p-p^*|
\end{equation}  
Qualitatively this agrees with other type I critical solutions.
\section{Different Families of Initial Data}
\label{universal}
\begin{figure}[htbp]
\epsfxsize=3.5in
\centerline{\epsfbox[100 190 550 690]{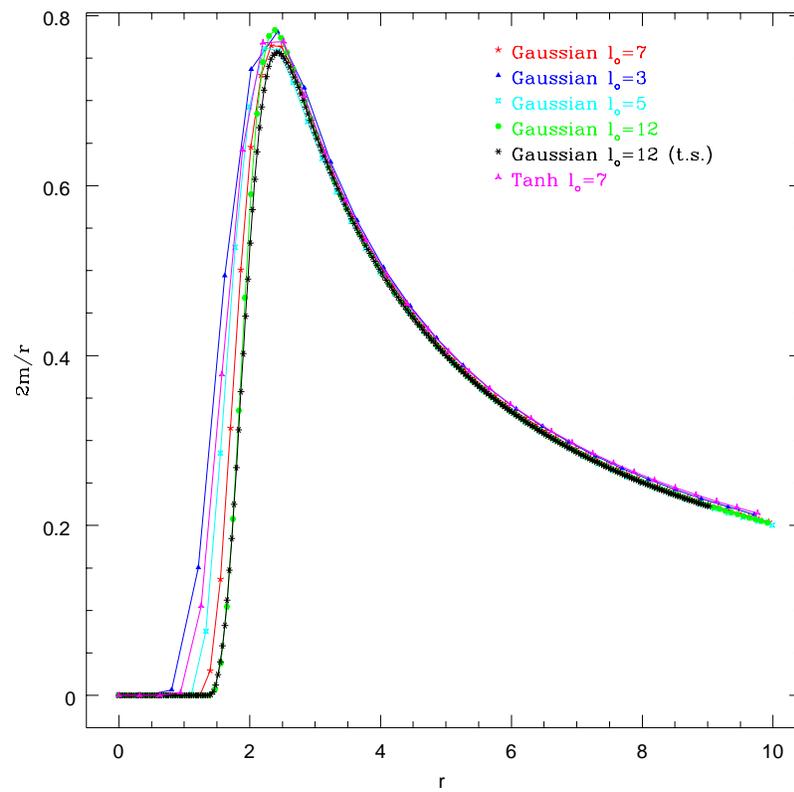}}
\caption[Critical solution for different families of initial data.]
{Critical solution for different families of initial data.}
\label{families}
\end{figure}

The results we showed in the previous section are for
Gaussian initial data given by equations [\ref{gaussian1}-
\ref{gaussian2}] with $p_r{}_o=0$ (``almost time symmetric" initial data)
and $l_o=7$.
In this section we show that similar results are found for other families of 
initial data. We have chosen initial data
with $p_r{}_o=-4$ and $l_o=3$, $l_o=5$, $l_o=7$, $l_o=12$, (with same $r_o=5$, $\Delta_{r}=1$, $\Delta_{\bar p_r}=2$,
$\Delta_{l}=2$ as before) and find that 
the critical solution for each of these cases appears to approach a static solution.
For smaller values of the angular momentum the mass in the critical solution 
gets concentrated closer to the origin, which makes it more difficult to resolve the
solution with a uniform  finite-difference grid. 
We also  have evolved an initial family with the following one particle distributions:
\begin{eqnarray}
\label{tanh1}
R(r)&\propto&\left(1-\tanh\left((r-r_o)/\Delta_r\right)^2 \right) \, \Theta (r) \\
P(p_r)&\propto&\left(1-\tanh\left((\bar p_r-{\bar p_r}{}_o)/\Delta_{pr}\right)\right)\\
\label{tanh2}
L(l)&\propto&\left(1-\tanh\left((l-l_o)/\Delta_l\right)^2 \right) \,\Theta (l)
\end{eqnarray}  
with $r_o=5$,$\Delta_r=1$, ${\bar p_r}{}_o=-4$, $\Delta_{\bar pr}=2$, $l_o=7$, 
$\Delta_l=2$. 
For this distribution we also find that the solution with small
$p-p^*$ ($p$ being the total rest mass of the cluster as before) spends some
time close to an apparently static solution.
In Figure \ref{families} we show profiles for $2M(r)/r$ for the different 
families, each separate profile being selected from the corresponding period of near-critical evolution.
Since different initial conditions set
different scales for the problem we have normalized the results such that the total ADM mass 
of the solutions to which they approach is $1$ at infinity (rescaling given by equations
(\ref{sca1}-\ref{sca2})).  
We can see that after normalization, the peak of $2M(r)/r$ is {\it roughly} 
at the same $r^*=2.3$. We can also see that the better a solution is resolved, 
the closer it conforms to the best resolved solution (the Gaussian with $l_o=7$ and ``almost 
time symmetric" initial data). This is an indication that the critical solution
might be universal although, we again need better resolution and more particles
to be able to give a definitive answer.

\begin{figure}[htbp] \epsfxsize=3.6in \centerline{\epsfbox[100 190 550 690]{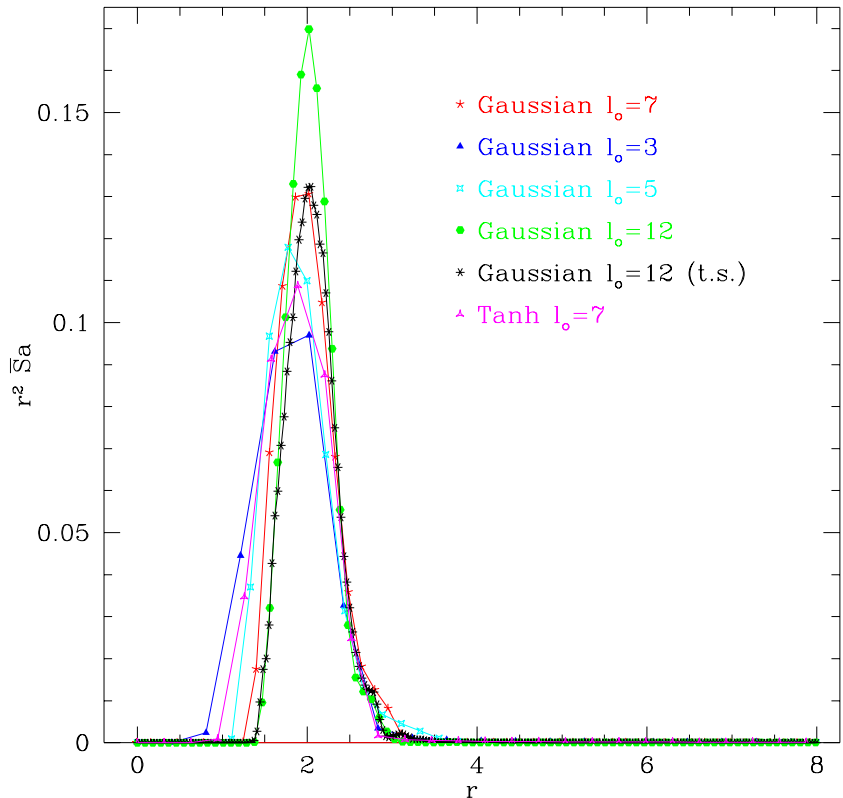}} \caption[$r^2 \bar S^a{}_a(r)$ for the different families during the critical regime.] {$r^2 \bar S^a{}_a(r)$ for the different families during the critical regime.} \label{r2sa} \end{figure}

We are also interested in seeing the importance of the distribution of angular momentum
on the critical solutions. In Figure \ref{r2sa} we show $r^2\, \bar S^a{}_a(r)$ where 
$\bar S^a{}_a(r)$ is the function defined by equations (\ref{inter}) and (\ref{sr_sa}) 
for the different families of initial data during the critical regime. This function, 
$r^2 S^a{}_a(r)$, is a dimensionless quantity which measures the square of the angular momentum 
of the 
distribution of particles. We have rescaled the radial coordinate (and time) in the same way as in
Figure \ref{families}. 
We can see that there is no apparent agreement between the 
functions calculated from different families of initial data.
More work needs to be done in order to see what is the dependence of the critical solution
with respect the angular momentum distribution. Moreover, estimates for the truncation and statistical errors
for each calculation of the critical solutions must be calculated to properly compare the 
solutions.

\begin{figure}[htbp]
\epsfxsize=5.0in
\centerline{\epsfbox[35 312 520 647]{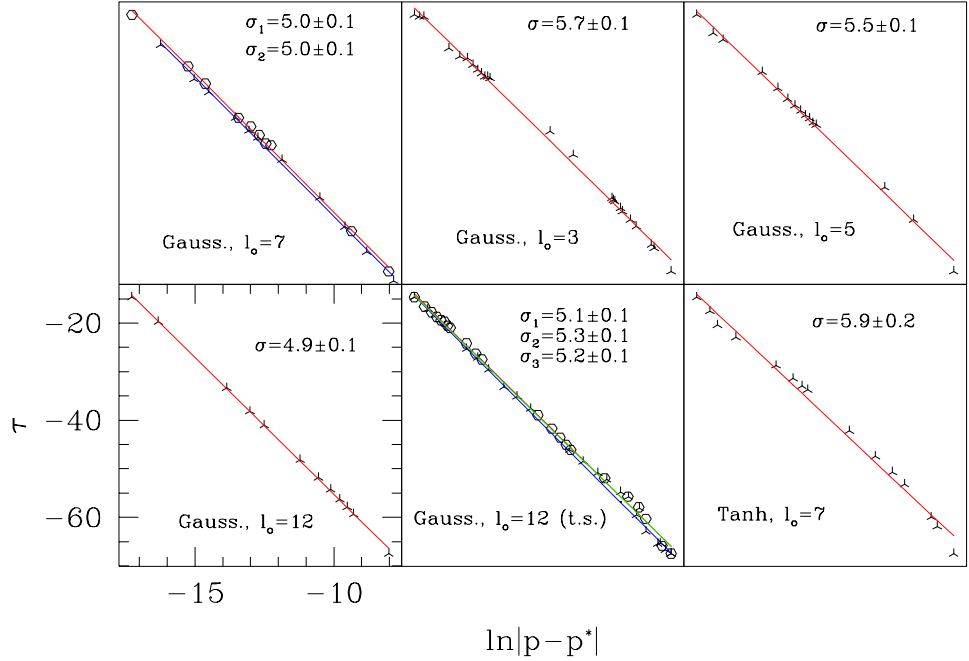}}
\caption[Scaling behaviour for different families of initial data .]
{Scaling behaviour for different families of initial data .}
\label{sigmasplot}
\end{figure}

We have also estimated $\sigma$ defined by 
\begin{equation}
\tau = -\sigma {\mathrm ln}|p-p^*|
\end{equation} for the different families.
In Figure \ref{sigmasplot} we show the scaling behaviour for the different families of initial data
that we have studied in this thesis.
The error for each value of $\sigma$ in the Figure is, as beforre, the standard deviation of the slope computed using a least squares 
fit. We have collected the values that we have obtained for $\sigma$ in Table \ref{sigmas}. 


\begin{table}[htbp]
\begin{center}
\begin{tabular}{| c | c |}\hline
Family & $\sigma$ \\ \hline \hline
\ \  Gaussian, \quad $l_o=7$ (set 1)\ \     & \ $5.0\pm0.2$ \ \\ \hline 
\ \  Gaussian,\quad $l_o=7$ (set 2)\ \  & \ $5.0\pm0.2$ \ \\ \hline 
\ \  Gaussian,\quad $l_o=3$ \ \   & \ $5.7\pm0.2$ \ \\ \hline 
\ \  Gaussian,\quad $l_o=5$ \ \   & \ $5.5\pm0.2$ \ \\ \hline 
\ \  Gaussian,\quad $l_o=12$ \ \   & \ $4.9\pm0.2$ \ \\ \hline 
\ \  Gaussian,\quad $l_o=12$ t.s. (set 1) \ \   & \ $5.1\pm0.2$ \  \\ \hline  
\ \  Gaussian,\quad $l_o=12$ t.s. (set 2) \ \   & \ $5.3\pm0.2$ \ \\ \hline 
\ \  Gaussian,\quad $l_o=12$ t.s. (set 3) \ \   & \ $5.2\pm0.2$ \ \\ \hline 
Tanh,\quad $l_o=7$   & $5.9\pm0.2$ \\ \hline
\end{tabular}
\end{center}
\caption[Values of time scaling exponent for the different families (t.s. stands for "almost time symmetric"). The errors above are assumed to be the same as 
the statistical error for the t.s. family (see equation (\ref{scal_eqn})).]
{Values of time scaling exponent for the different families (t.s. stands for "almost time symmetric"). The errors above are assumed to be the same as the statistical error for the t.s. family (see equation (\ref{scal_eqn})).}
\label{sigmas}
\end{table}


\section{Conclusions and Future Work}
We have studied critical behavior at the threshold of black hole formation for collisionless
matter.
Our results indicate that, for families with non-zero angular momentua, the critical solution could be static 
with non-zero radial particle momenta. We have found evidence for a life-time scaling law which is to be expected 
for Type I critical solutions. 
In order to answer all these questions more rigorously we would need  
some way to increase our resolution where it is needed. Some kind of adaptive
code (such as the ones used in finite difference studies \cite{Choptuik:1993}) would be useful.

It may be that the development of a finite difference code to solve the \Vlasov\ equation directly in phase space
would be the best way to attack this problem. This would allow us to incorporate the adaptivity 
that we need, as well as providing us with better-understood convergence properties.

   \bibliography{msc}
   \begin{appendix}
   \chapter{Summary of Equations in Maximal-Areal Coordinates}
\section{Field equations}
\begin{eqnarray}
\frac{a'}{a}&=&\frac{1-a^2}{2r}+\frac{3}{2}ra^2 {K^\theta_\theta}^2 + 4\pi
a r \bar \rho\\
{K^\theta_\theta}'&=&-\frac{3}{r}K^\theta_\theta-4\pi \frac{\bar j_r}{a}\\
\alpha''&=&\alpha'(\frac{a'}{a}-\frac{2}{r})+\frac{2\alpha}{r^2}(a^2-1+2r\frac{a'}{a})+4\pi
a \alpha (\frac{\bar S^r_r}{a^2}+\bar S^\theta_\theta-3 \bar \rho)\\
\beta&=&\alpha r K^\theta _\theta
\end{eqnarray}
where
\begin{eqnarray}
\bar \rho &=&\frac{1}{4\pi\Delta r } \sum_{i=1}^N \frac{\bar p^t_i}{r^2_i} W(r-r_i)\\
\bar S^r_r &=&\frac{1}{4\pi\Delta r } \sum_{i=1}^N \frac{{{p_r}_i}^2}{\bar p^t_ir^2_i} W(r-r_i)\\
\bar S^\theta_\theta &=&\frac{1}{4\pi\Delta r} \sum_{i=1}^N \frac{l^2_i}{r^4_i \bar p^t_i} W(r-r_i)\\
\bar \j_r &=&\frac{1}{4\pi\Delta r} \sum_{i=1}^N \frac{{p_r}_i}{r^2_i} W(r-r_i)
\end{eqnarray}
and
\begin{equation}
W(x-x_i)= \left\{ \begin{array}
{r@{\quad:\quad}l}
1-|x-x_i|/h & |x-x_i| \le h \\ 0 & {\mathrm otherwise} \end{array} \right. 
\end{equation}
\section{Evolution Equations}
\begin{eqnarray}
\frac{d p_r}{d t} &=&-\frac{\partial\alpha}{\partial r} \bar p^t+
\frac{\partial \beta}{\partial r} p_r +
\frac{\alpha}{a^3}\frac{\partial a}{\partial r}\frac{{p_r}^2}{\bar p^t}+
\frac{l^2 \alpha}{\bar p^t r^3}\\
\frac{d r}{d t}&=&\frac{\alpha p_r}{a^2 \bar p^t}-\beta\\
\bar p^t& =& \sqrt{m^2+\frac{{p_r}^2}{a^2}+\frac{l^2}{r^2}}\\
\frac{dp^\theta}{dt}&=&-\frac{2}{r}\left(\frac{\alpha p_r}{a^2 \bar p^t}
- \beta \right) p^\theta + \frac{1}{\tan{\theta}} \frac{\alpha}{\bar p^t}
\left(\frac{l^2}{r^4}-{p^\theta}^2\right)\\
\frac{d \theta}{dt}&=&\frac{\alpha p^\theta}{\bar p^t}\\
{p^\phi}^2&=&\frac{1}{\sin^2{\theta}}\left( \frac{l^2}{r^4} - {p^\theta}^2\right)\\
\frac{d \phi}{dt}&=&\frac{\alpha p^\phi}{\bar p^t}
\end{eqnarray} 


\chapter{Direct  Approach To Solutions of Einstein-Vlasov System}
\section{Field Equations}
The distribution function is defined by:
\begin{equation}
f(t,r,p^r,l)=\frac{dN}{dV},
\end{equation}
where $N$ is the number of particles and $V$ is the volume in phase space.
Then the stress energy tensor can be calculated via:
\begin{equation}
T^{\mu\nu}=\int\, p^\mu\, p^\nu\, f(t,r,p^r,l)\,\frac{1}{m}\,dV_p
\end{equation}
where the volume element in momentum space, $V_p$, is restricted by
\begin{equation}
-m^2=g_{\alpha\beta}\,p^{\alpha}\,p^{\beta}
\end{equation}
to the following expression (see \cite{Shapiro:1989}):
\begin{equation}
dV_p=m \frac{dp_r\, dp_\theta\, dp_\phi}{p^t\, \sqrt{-g}}
\end{equation}
We can introduce momentum coordinates adapted to the symmetry:
\begin{eqnarray}
l^2&=&p_\theta^2 + \frac{p_\phi^2}{\sin^2{\theta}}\\
\psi&=&\tan^{-1}{\frac{p_\theta \sin{\theta}}{p_\phi}}
\end{eqnarray}
and $\bar p^t$ given, as before, by:
\begin{equation}
\bar p^t=\sqrt{m^2+\frac{p_r^2}{a^2}+\frac{l^2}{r^2}}
\end{equation}
The volume element $dV_p$ in these variables is:
\begin{equation}
dV_p=m \frac{dp_rdl^2d\psi}{2 a r^2 \bar p^t}
\end{equation}
The stress-energy tensor components are:
\begin{eqnarray}
T^t_t&=&\frac{-\pi}{a^2r^2}\int\bar p^t f dp_r dl^2 +
\frac{\beta \pi }{\alpha a^2 r^2}\int p_r f dp_r dl^2\\
T^t_r&=&\frac{\pi}{a^2 r^2 \alpha} \int p_r f dp_r dl^2\\
T^r_r&=&\frac{-\beta \pi}{\alpha a^2 r^2} \int p_r f dp_r dl^2 +
\frac{\pi}{a^4 r^2} \int \frac{{p_r}^2}{\bar p^t} f dp_r dl^2\\
T^\theta_\theta=T^\phi_\phi&=&\frac{1}{2}\frac{\pi}{a^2r^4}
\int \frac{l^2}{\bar p^t} dp_r dl^2
\end{eqnarray}
We are able then to construct the source terms:
\begin{eqnarray}
\rho&=&\frac{\pi}{a r^2}
\int{f(t,r,p_r,l)\sqrt{m^2+\frac{p_r^2}{a^2}+\frac{l^2}{r^2}} dp_rdl^2}\\
j_r&=&\frac{\pi}{ar^2}\int{f(t,r,p_r,l) p_r dp_r dl^2}\\
S^r_r&=&\frac{\pi}{a^3r^2}\int{\frac{p_r^2 f(t,r,p_r,l)}{\sqrt{m^2+p_r^2/a^2+l^2/r^2}}
dp_r dl^2}\\
S^\theta_\theta&=&\frac{\pi}{a r^4}\int\frac{l^2
f(t,r,p_r,l)}{\sqrt{m^2+p_r^2/a^2+l^2/r^2}} dl^2 dp_r
\end{eqnarray}
\section{\Vlasov\ Equation} 
If the system is collisionless, we have
Liouville's theorem concerning the conservation of the volume in phase space, $V$,
and we
also know that the particle number $N$ is a conserved quantity.
Therefore we can write the conservation of the distribution function,
$f=N/V$, as:
\begin{equation}
\frac{d f}{d \tau}=0
\end{equation}
This equation is known as the ``collisionless Boltzmann's equation" or the ``kinetic
equation". Expanding the total derivative we have:
\begin{equation}
\frac{\partial x^\alpha}{\partial \tau}\frac{\partial f}{\partial x^\alpha} +\\
\frac{\partial p_\alpha}{\partial \tau}\frac{\partial f}{\partial p_\alpha} = 0
\end{equation}
where $x^\alpha$ are the space-time variables, and $p^\alpha$ are the momentum
variables.
In spherical symmetry this equation reduces to:
\begin{equation}
\frac{\partial f}{\partial t}+\\
\frac{\partial r}{\partial t}\frac{\partial f}{\partial r}+\\
\frac{\partial p_r}{\partial t}\frac{\partial f}{\partial p_r} = 0
\end{equation}
where $\partial r/ \partial t$ and $\partial p_r/\partial t$ 
are given by equations (\ref{evol1}) and (\ref{evol2}). This gives:
\begin{equation}
\frac{\partial f}{\partial t}+\\
(\frac{\alpha p_r}{a^2 \bar p^t}-\beta)\frac{\partial f}{\partial r}+\\
(-\frac{\partial\alpha}{\partial r} \bar p^t+
\frac{\partial \beta}{\partial r} p_r +
\frac{\alpha}{a^3}\frac{\partial a}{\partial r}\frac{{p_r}^2}{\bar p^t}+
\frac{\alpha l^2}{\bar p^t r^3})\frac{\partial f}{\partial p_r} = 0
\end{equation}
\chapter{Summary of Equations in Polar-Areal Gauge Coordinates}
In this section we present the equations for a second choice of coordinates.
Specifically we again adopt areal radial coordinates (implying that the
proper surface area of the spheres centered at $r=0$ and with radius $r$ will be $4\pi r^2$),
but fix the time coordinate using polar slicing. This slicing condition is implemented by
demanding:
\begin{equation}
{\mathrm Tr} K \equiv K^i{}_i=K^r{}_r 
\end{equation}
which implies $K^\theta{}_\theta=K^\phi{}_\phi=\dot K^\theta{}_\theta=\dot K^\phi{}_\phi=0$.
Using expression (\ref{ktt}) we can express $K^\theta{}_\theta$ as:
\begin{equation}
K^\theta{}_\theta=\frac{1}{rb\alpha}(\beta(rb)^\prime-r\dot b)=0,
\end{equation}
we have:
\begin{equation}
\beta(rb)^\prime=r\dot b
\end{equation}
Choosing areal coordinates, $b=1$, $\dot b=0$, the above
equation implies:
\begin{equation}
\beta=0.
\end{equation}
The resulting metric is:
\begin{equation}
ds^2=-\alpha^2(t,r)dt^2+a^2(t,r)dr^2+r^2d\Omega
\end{equation}
we then have the following geometrical quantities:
\begin{equation}
\begin{array}{c}
R=2\left[-2\left(1/a\right)^\prime+
  a/r\left(1-1/a^2\right)\right]/(ar)\\
\\
\begin{array}{ll}
R^r{}_r=2a^\prime/\left(r a^3\right)&
R^\theta{}_\theta=R^\phi{}_\phi=1/\left(ar^2\right)\left(a-\left(r/a\right)^\prime\right)\\
\\
K^r{}_r=K=-\dot a/\left(a \alpha\right)&
K^\theta{}_\theta=K^\phi{}_\phi=0\\
\\
{K^i}{}_{r|i}=-\left(\dot a/\left(a \alpha\right)\right)^\prime
           -2\dot a / \left(r a \alpha\right)&
K_{|r}={K^r{}_r}_{,r}=-\left(\dot a/\left(a \alpha\right) \right)\\
\\
{\alpha_{|r}}^{|r}=1/a\left(\alpha^\prime/a\right)^\prime&
{\alpha_{|\theta}}^{|\theta}={\alpha_{|\phi}}^{|\phi}=\alpha^\prime/\left(a^2r\right)\end{array}
\end{array}
\end{equation}
Briefly we summarize the resulting equations of motion:

{\it Momentum Constraint:}
\begin{equation}
\dot a= -4 \pi r a \alpha j_r
\end{equation}

{\it Hamiltonian Constraint:}
\begin{equation}
a^\prime=4\pi r a^3\rho-\frac{a}{2r}(a^2-1)
\end{equation}

{\it Evolution of the 3-metric}
\begin{equation}
\dot g_{ij}=-2\alpha g_{ik}K^k_j
\end{equation}

{\it Evolution of the extrinsic curvature}
\begin{eqnarray}
\dot K^r{}_r&=&-\frac{1}{a}\left(\frac{\alpha^\prime}{a}\right)^\prime+
            \frac{2\alpha a^\prime}{ra^4}+4\pi\alpha(S-\rho)-8\pi \alpha S^r{}_r\\\dot K^\theta{}_\theta&=&0=
-\frac{\alpha^\prime}{a^2r}+\frac{\alpha}{ar^2}\left(a-\frac{1}{a}+r\frac{a^\prime}{a^2}\right)
+4\pi \alpha(S-\rho)-8\pi \alpha S^\theta{}_\theta\\
\dot K^\phi{}_\phi&=&0=
-\frac{\alpha^\prime}{a^2r}+\frac{\alpha}{ar^2}\left(a-\frac{1}{a}+r\frac{a^\prime}{a^2}\right)+4\pi \alpha(S-\rho)-8\pi \alpha S^\phi{}_\phi
\end{eqnarray}

where the two last equations are zero by the choice of slicing.

   \chapter{Pseudo-code} 
\label{pseudo}
\begin{figure}[h]
\small
{\bf routine} vlasov \\
\hspace*{0.1cm} \# $r_i$ := position of the i-th particle \\
\hspace*{0.1cm} \# $[p_\mu]_i$ := momentum of i-th the particle \\ 
\hspace*{0.1cm} \# $[T^\mu_\nu]_j$ := $T^\mu_\nu$ component on the j grid point \\
\hspace*{0.1cm} \# $a_j$, $\alpha_j$ and $\beta_j$ are the metric coefficients \\
\hspace*{0.1cm} \# $M_j$ := mass aspect function \\
\hspace*{0.1cm} \# Generate the initial positions $r_i$ and velocities $[p_\mu]_i$ for the particles \\
\hspace*{0.1cm} {\bf read} parameters of the distribution function \\
\hspace*{0.1cm} {\bf for} $i=1...N_{particles}$ \\  
\hspace*{0.5cm} Generate $r_i$ and $[p_\mu]_i$ using Monte Carlo for a given distribution function\\
\hspace*{0.1cm} {\bf end for} \\  
\hspace*{0.1cm} t = 0 \\  
\hspace*{0.1cm} \# Time loop \\  
\hspace*{0.1cm} {\bf while} $t < t_{max}$  \\  
\hspace*{0.75cm} {\bf for} $j=1...N_{grid points}$ \\  
\hspace*{1.25cm} Calculate $[T^\mu_\nu]_j$ from $r_i$ and $[p_\mu]_i$\\
\hspace*{0.75cm} {\bf end for} \\  
\hspace*{0.75cm} Solve for $a_j$, $\beta_j$ and $\alpha_j$ on the mesh \\  
\hspace*{0.75cm} \# Check apparent horizon formation \\  
\hspace*{0.75cm} {\bf if} $2M_j/r_j = 1$ \ {\bf stop}\\  
\hspace*{0.75cm} Ouput $a_j$, $\beta_j$ and $\alpha_j$ \\  
\hspace*{0.75cm} {\bf for} $i = 1...N_{particles}$ \\  
\hspace*{1.25cm} Calculate $[\Gamma^\mu_{\nu \eta}]_i$ at $t=t+1/2\ dt$ 
from $a_j$, $\beta_j$ and $\alpha_j$ \\
\hspace*{0.75cm} {\bf end for}\\   
\hspace*{0.75cm} t = t + dt\\
\hspace*{0.75cm} {\bf for} $i = 1...N_{particles}$ \\  
\hspace*{1.25cm} Integrate geodesic equations to find new $r_i$ and $[p_\mu]_i$\\
\hspace*{1.25cm} {\bf if} $r_i>r_{max}$  Kill i-th particle, $N_{particles}=N_{particles}-1$\\
\hspace*{1.25cm} {\bf if} $r_i<0$  \ \ \ Reflect the particle\\
\hspace*{1.25cm} {\bf if} $N_{particles}=0$  \ \ \ {\bf stop}\\
\hspace*{0.75cm} {\bf end for}  \\  
\hspace*{0.1cm} {\bf end while} \\  
{\bf end routine} \\  
\end{figure}

   \end{appendix}	
\end{document}